\documentclass[prb, onecolumn]{revtex4}
\usepackage{ae, aecompl}
\usepackage{helvet}
\usepackage{epsfig}
\usepackage{dcolumn}
\usepackage{color}
\usepackage[LGR, T1]{fontenc}
\usepackage[latin9]{inputenc}
\usepackage{amssymb}
\usepackage{amsmath}
\usepackage{mwe, tikz}
\usepackage{esint}
\usepackage{longtable}
\usepackage{url}
\usepackage{etoolbox}
\usepackage{multirow}
\usepackage{soul}
\usepackage{rotating}

\newcommand{\beginsupplement}{%
 \setcounter{table}{0}
 \renewcommand{\thetable}{S\arabic{table}}%
 \setcounter{figure}{0}
 \renewcommand{\thefigure}{S\arabic{figure}}%
 }

\renewcommand\footnotemark{}

\makeatletter
\def\blfootnote{\gdef\@thefnmark{}\@footnotetext}
\makeatother

\makeatletter
\renewcommand{\fnum@figure}{\textbf{Figure \thefigure}}
\makeatother

\DeclareFontEncoding{LGR}{}{}
\DeclareTextSymbol{\~}{LGR}{126}

\begin{document}
\bibliographystyle{apsrev}

\title{Transition from Fireball to Poynting-flux-dominated Outflow in the Three-Episode GRB 160625B}
\author{B.-B. Zhang$^{1, 2, 3}$, 
B. Zhang$^{4,5,6}$, 
A. J. Castro-Tirado$^{1, 7}$, 
Z. G. Dai$^{2, 8}$, 
P.-H. T. Tam$^{9}$, 
X.-Y. Wang$^{2, 8}$, 
Y.-D. Hu$^{1, 10}$, 
S. Karpov$^{11, 12}$, 
A. Pozanenko$^{13,14,15}$, 
F.-W. Zhang$^{16}$, 
E. Mazaeva$^{13}$, 
P. Minaev$^{13}$, 
A. Volnova$^{13}$,
S. Oates$^{17}$, 
H. Gao$^{18}$, 
X.-F. Wu$^{19, 20, 21}$, 
L. Shao$^{22,19}$, 
Q.-W. Tang$^{23 }$, 
G. Beskin$^{11, 12}$, 
A. Biryukov$^{12, 24}$, 
S. Bondar$^{25}$, 
E. Ivanov$^{25}$, 
E. Katkova$^{25}$, 
N. Orekhova$^{25}$, 
A. Perkov$^{25}$, 
V. Sasyuk$^{12}$, 
L. Mankiewicz$^{26}$, 
A. F. \.Zarnecki$^{27}$, 
A. Cwiek$^{28}$, 
R. Opiela$^{26}$, 
A. Zadro\.zny$^{28}$, 
R. Aptekar$^{29}$, 
D. Frederiks$^{29}$, 
D. Svinkin$^{29}$, 
A. Kusakin$^{30}$, 
R. Inasaridze$^{31}$, 
O. Burhonov$^{32}$, 
V. Rumyantsev$^{33}$, 
E. Klunko$^{34}$, 
A. Moskvitin$^{11}$, 
T. Fatkhullin$^{11}$, 
V. V. Sokolov$^{11}$, 
A. F. Valeev$^{11, 12}$, 
S. Jeong$^{1, 35}$, 
I. H. Park$^{35}$, 
M. D. Caballero-Garc\'ia$^{36}$, 
R. Cunniffe$^{1}$, 
J. C. Tello$^{1}$, 
P. Ferrero$^{1}$, 
S. B. Pandey$^{37}$, 
M. Jel\'inek$^{36}$,
F. K. Peng$^{38}$,
R. S{\'a}nchez-Ram{\'{\i}}rez$^{1}$,
A. Castell\'on$^{39}$}%

\begin{abstract}

\end{abstract}

 \maketitle
\textbf{
The ejecta composition is an open question in gamma-ray bursts (GRB) physics \cite{KumarZhang2015}. Some GRBs possess a quasi-thermal spectral component in the time-resolved spectral analysis \citep{Ryde2010}, suggesting a hot fireball origin. Others show a featureless non-thermal spectrum 
known as the ``Band" function \citep{Band1993,Preece2014,Zhang2016}, consistent with a synchrotron radiation origin \citep{Uhm2014, Zhang2016} and suggesting that the jet is Poynting-flux-dominated at the central engine and likely in the emission region as well \citep{Hascoet12,ZhangPeer2009}. There are also bursts showing a sub-dominant thermal component and a dominant synchrotron component \cite{Axelsson2012},
suggesting a likely hybrid jet composition \cite{GaoZhang2015}. Here we report an extraordinarily bright GRB 160625B, simultaneously observed in gamma-rays and optical wavelengths, whose prompt emission consists of three isolated episodes separated by long quiescent intervals, with the durations of each ``sub-burst" being $\sim$ 0.8 s, 35 s, and 212 s, respectively. Its high brightness (with isotropic peak luminosity L$_{\rm p, iso}\sim 4\times 10^{53}$ erg/s) 
allows us to conduct detailed time-resolved spectral analysis in each episode, from precursor to main burst and to extended emission. The spectral properties of the first two sub-bursts are distinctly different, allowing us to observe the transition from thermal to non-thermal radiation between well-separated emission episodes within a single GRB. Such a transition is a clear indication of the change of jet composition from a fireball to a Poynting-flux-dominated jet.}

\blfootnote{\footnotesize{
$^{1}$Instituto de Astrof\'isica de Andaluc\'ia (IAA-CSIC), P.O. Box 03004, E-18080 Granada, Spain; zhang.grb@gmail.com;
$^{2}$School of Astronomy and Space Science, Nanjing University, Nanjing 210093, China;
$^{3}$Scientist Support LLC, Madison, AL 35758, USA;
$^{4}$Department of Physics and Astronomy, University of Nevada, Las Vegas, NV 89154, USA; zhang@physics.unlv.edu
$^{5}$ Department of Astronomy, School of Physics, Peking University, Beijing 100871, China;
$^{6}$ Kavli Institute for Astron. Astrophys., Peking University, Beijing 100871;
$^{7}$Departamento de Ingenier\'ia de Sistemas y Autom\'atica, Escuela de Ingenier\'ias, Universidad de M\'alaga, C\/. Dr. Ortiz Ramos s\/n, 29071 M\'alaga, Spain; 
$^{8}$Key laboratory of Modern Astron. Astrophys. (Nanjing University), Ministry of Education, Nanjing 210093, China;
$^{9}$School of Physics and Astronomy, Sun Yat-Sen University, Zhuhai 519082, China;
$^{10}$Universidad de Granada, Facultad de Ciencias Campus Fuentenueva s/n E-18071 Granada, Spain;
$^{11}$Special Astrophysical Observatory of Russian Academy of Sciences, Nizhniy Arkhyz, Russia;
$^{12}$Kazan Federal University, Kazan, Russia;
$^{13}$Space Research Institute of the Russian Academy of Sciences, 117997, Russia, Moscow, Profsoyuzanaya, 84/32;
$^{14}$National Research Nuclear University MEPhI (Moscow Engineering Physics Institute), 115409, Russia, Moscow, Kashirskoe shosse, 31;
$^{15}$ National Research University Higher School of Economics, 101000, 20 Myasnitskaya STR, Moscow, Russia;
$^{16}$College of Science, Guilin University of Technology, Guilin 541004, China;
$^{17}$Department of Physics, University of Warwick, Coventry, CV4 7AL, UK;
$^{18}$Department of Astronomy, Beijing Normal University, Beijing 100875, China;
$^{19}$Purple Mountain Observatory, Chinese Academy of Sciences, Nanjing 210008, China;
$^{20}$School of Astronomy and Space Science, University of Science and Technology of China, Hefei, Anhui 230026, China;
$^{21}$Joint Center for Particle, Nuclear Physics and Cosmology, Nanjing University-Purple Mountain Observatory, Nanjing 210008, China;
$^{22}$Department of Space Sciences and Astronomy, Hebei Normal University, Shijiazhuang 050024, China;
$^{23}$Department of Physics, Nanchang University, Nanchang 330031, China;
$^{24}$Moscow State University, Moscow, Russia;
$^{25}$Research and Production Corporation ``Precision Systems and Instruments'', Nizhniy Arkhyz, Russia;
$^{26}$Center for Theoretical Physics PAS, Al. Lotnikow 32/46, 02-668 Warsaw, Poland;
$^{27}$Faculty of Physics, University of Warsaw, Pasteura 5, 02-093 Warsaw, Poland;
$^{28}$National Centre for Nuclear Research, Ho\.za 69, 00-681 Warsaw, Poland;
$^{29}$Ioffe Institue, Politekhnicheskaya ul., 26, St. Petersburg, 194021, Russia;
$^{30}$Fesenkov Astrophysical Institute, Kazakhstan, Almaty, 050020, 23;
$^{31}$Kharadze Abastumani Astrophysical Observatory, Ilia State University, Georgia, Tbilisi, 0162, Kakutsa Cholokashvili Ave 3/5;
$^{32}$Ulugh Beg Astronomical Institute, Uzbekistan, 100052, Tashkent, Astronomicheskaya st., 33;
$^{33}$Crimean Astrophysical Observatory, 298409, Nauchny, Crimea;
$^{34}$Institute of Solar-Terrestrial Physics, 664033, Russia, Irkutsk p/o box 291; Lermontov st., 126a
$^{35}$Department of Physics, Sungkyunkwan University (SKKU), Suwon, 440-746 Korea;
$^{36}$Astronomical Institute of the Academy of Sciences, Bocni II 1401, CZ-141 00 Praha 4 -- Sporilov, Czech Republic;
$^{37}$ARIES, Manora Peak, Nainital, India;
$^{38}$School of Physics and Electronic Science, Guizhou Normal University, Guiyang, Guizhou 550001, People's Republic of China;
$^{39}$Departamento de Algebra, Geometr\'ia y Topolog\'ia, Facultad de Ciencias, Universidad de M\'alaga, C\/. Juan L\'opez Pe\~nalver, s\/n, 29071 M\'alaga, Spain
}}

\begin{figure*}

\begin{tabular}{lll}
& & \multirow{ 3}{*}{\includegraphics[scale=1.3]{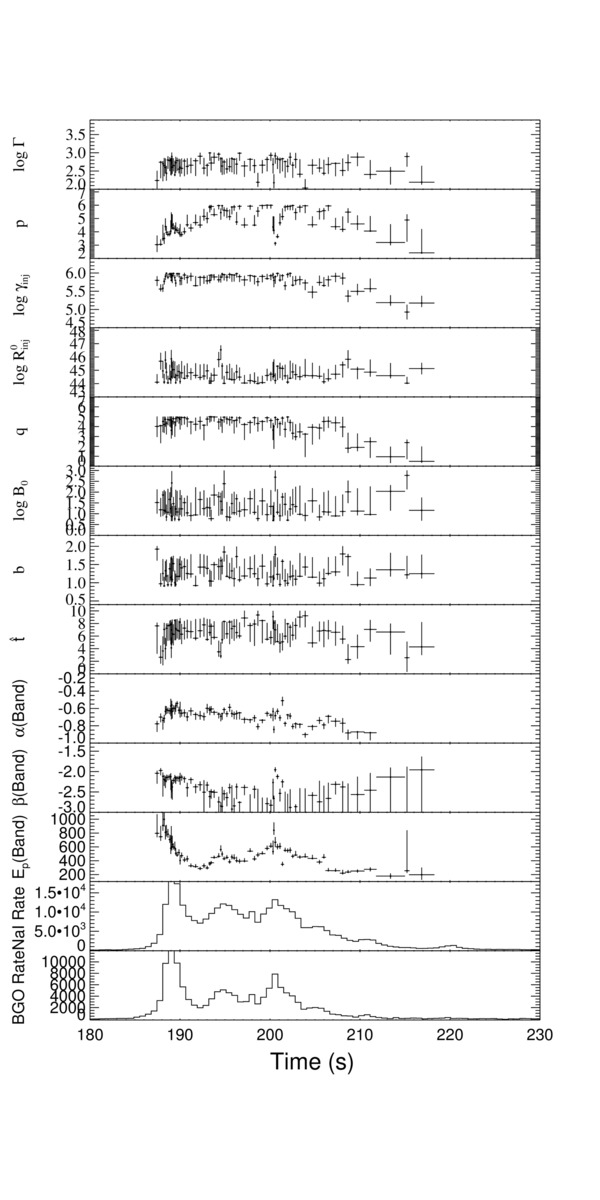}} \\
 & & \\
 \multicolumn{2}{c}{\includegraphics[keepaspectratio, clip, width=0.59\textwidth]{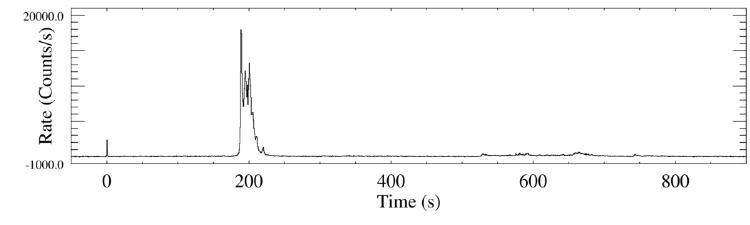}}& \\

\includegraphics[keepaspectratio, clip, width=0.33\textwidth]{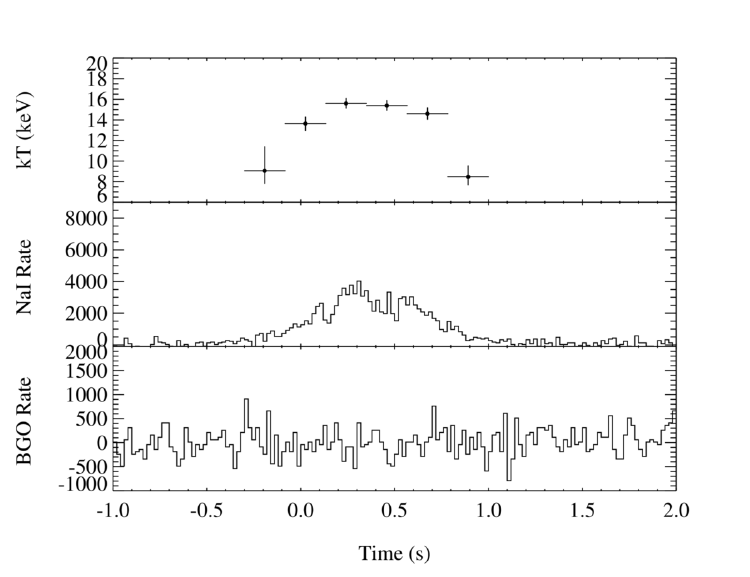} & \includegraphics[keepaspectratio, clip, width=0.33\textwidth]{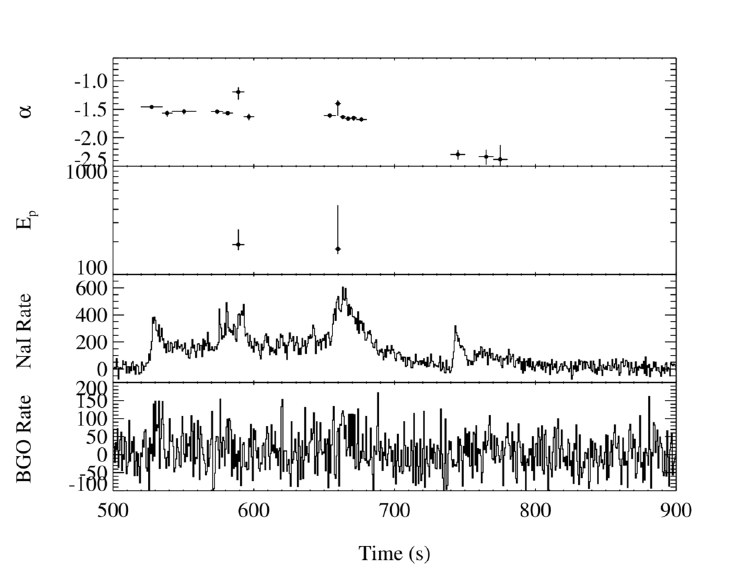} & \\

\end{tabular}

\caption{{\bf {\it Fermi}/GBM Light curve and spectral evolution of GRB 160625B}. \textbf{a}: prompt {\it Fermi}/GBM light curve of GRB 160625B. \textbf{b,c,d}: Spectral evolution of three sub-bursts. In each of these plots, the bottom two panels show the NaI and BGO light curves. The evolution of model parameters are displayed in the higher panels. The spectral models are blackbody for the 1st sub-burst, Band function and synchrotron model for the 2nd sub-burst.  For the 3rd sub-burst, a power law model is applied except two slices for which a power-law model with exponential cutoff is applied and $E_p$ is constrained. All the error bars represents 1-$\sigma$ uncertainties. 
}
\label{fig:totallc}
\end{figure*}

\begin{figure*}

\begin{tabular}{cc}
\includegraphics[keepaspectratio, clip, width=0.45\textwidth]{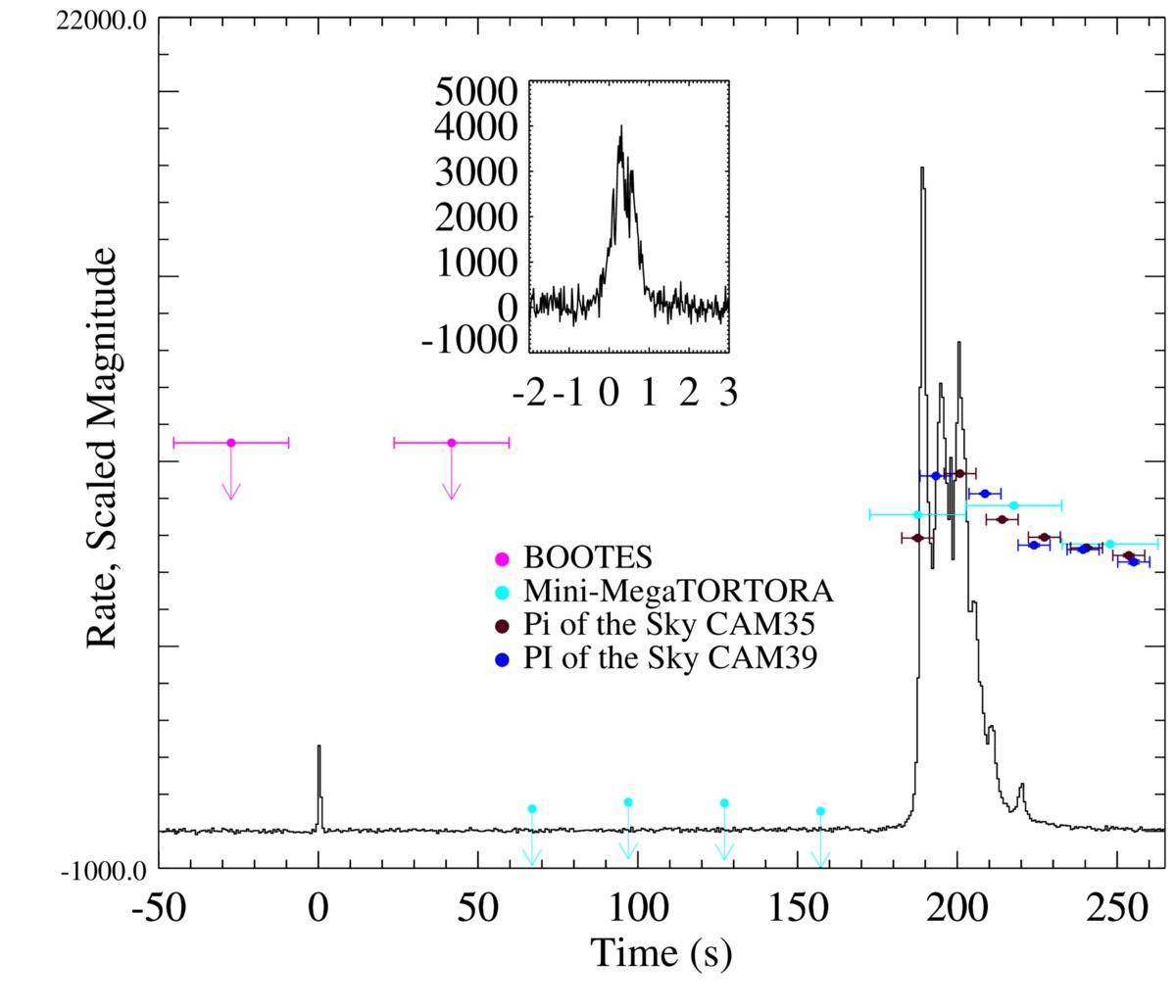} & \includegraphics[keepaspectratio, clip, width=0.45\textwidth]{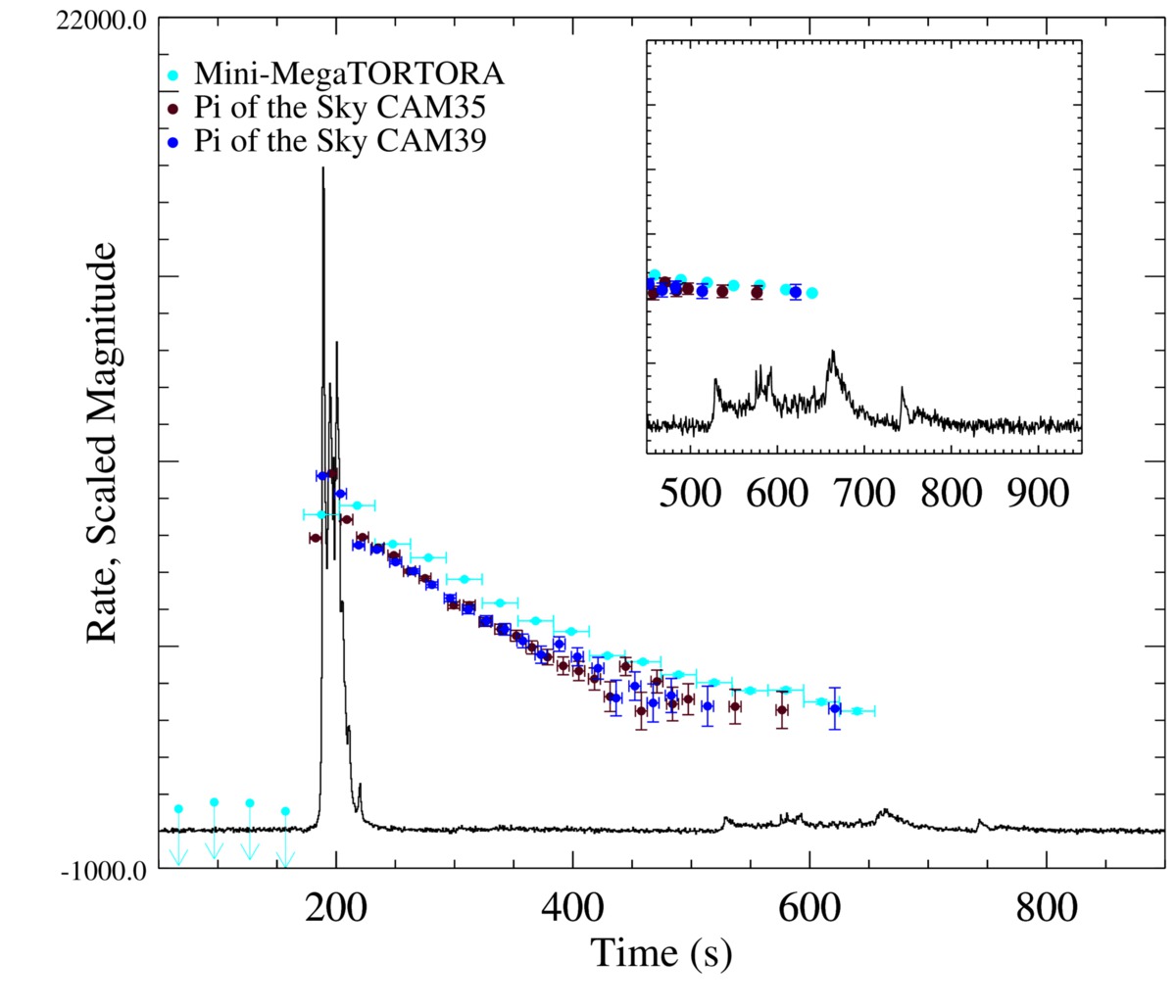} \\
 \multicolumn{2}{c}{\includegraphics[keepaspectratio, clip, width=0.89\textwidth]{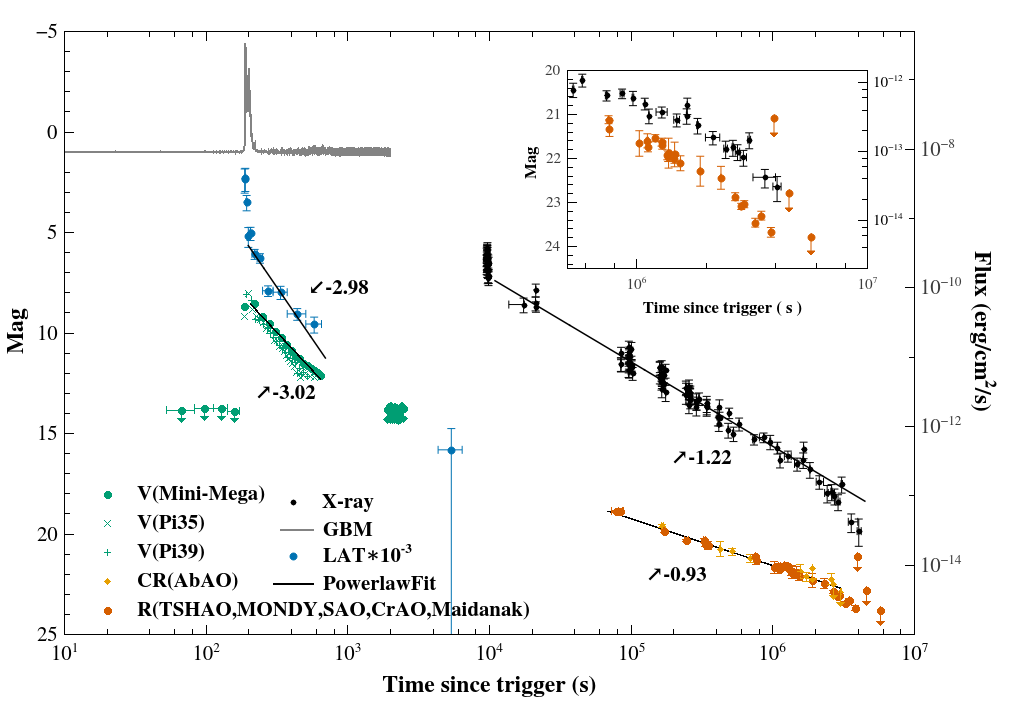} }\\

\end{tabular}

\caption{{\bf Multi-wavelength Observaiton of GRB 160625B.} {\bf a:} Optical Observations of the BOOTES-1, BOOTES-2, Pi of the Sky and MiniMega telescopes during the prompt emission in comparison with the GRB 160625B $\gamma$-ray light curve during the first and second sub-bursts.  {\bf b:} Same as a but during the second and third sub-bursts.  {\bf c:} Long-term multi-wavelength light curves. Inset shows the zoomed-in late-time optical and X-ray light curves.}
\label{fig:longterm}
\end{figure*}

GRB 160625B triggered Gamma-Ray Monitor (GBM; 8 keV - 40 MeV) on board the NASA {\it Fermi} Gamma-Ray Observatory twice at 22:40:16.28 UT and 22:43:24.82 on 25 June 2016 \cite{Troja2017}. The intense high-energy photon flux of GRB 160625B also triggered the Large Area Telescope (LAT) on board {\it Fermi}, Konus-{\it Wind}, {\it INTEGRAL}/SPI-ACS, RHESSI and CALET (attached to the {\it ISS}). The prompt emission light curves and spectral fitting results are shown in Figure \ref{fig:totallc}. Long term follow-up observations of the burst was carried out by {\it Swift} and several ground-based optical telescopes (see Supplementary Information for the details). The overall long term multi-wavelength afterglow light curves are presented in Figure \ref{fig:longterm}. A summary of the timing and spectral properties of the burst are listed in Table \ref{tab:sum}.

\begin{figure*}
\begin{tabular}{cc}
\includegraphics[keepaspectratio, clip, width=0.52\textwidth]{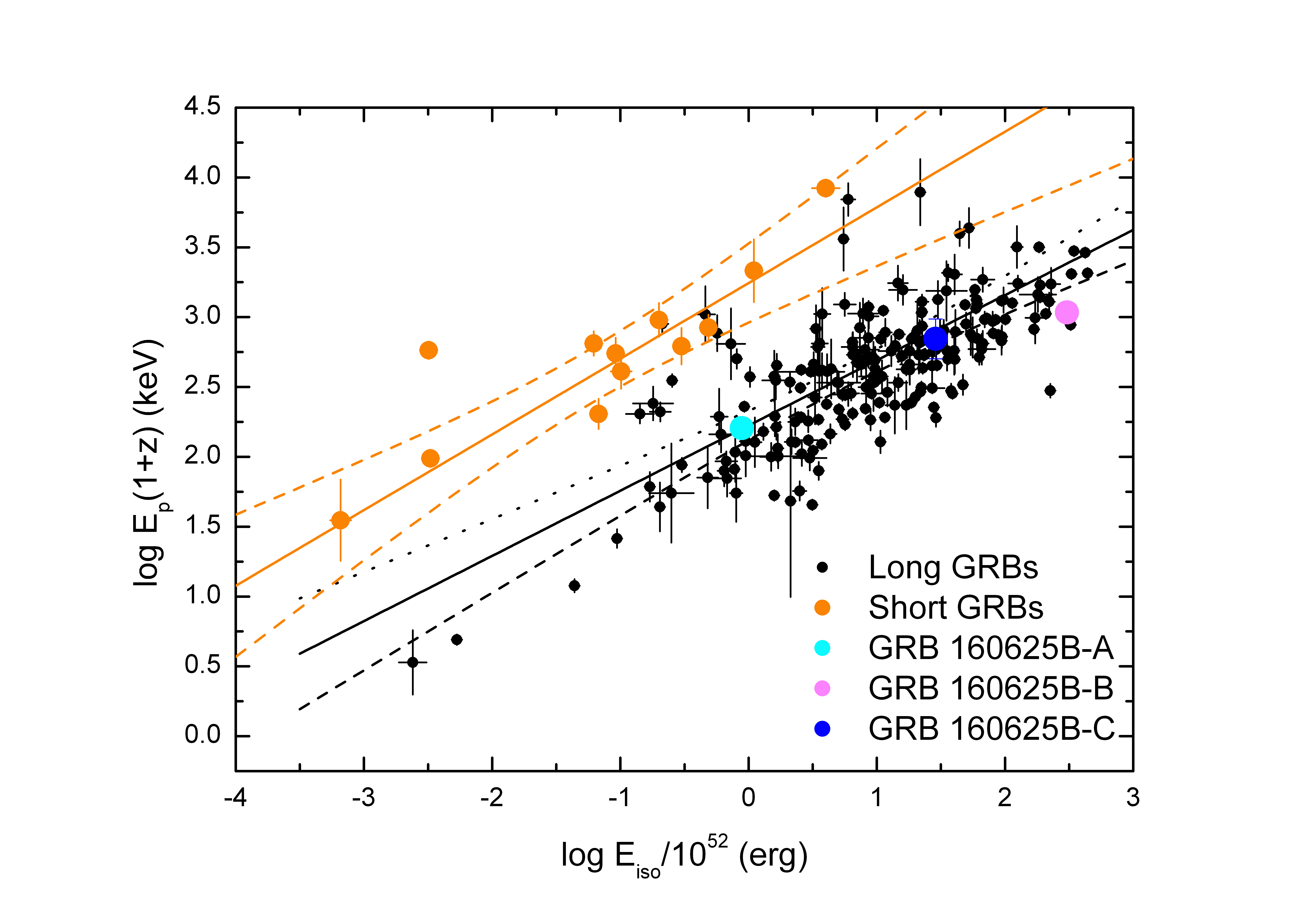} & \includegraphics[keepaspectratio, clip, width=0.52\textwidth]{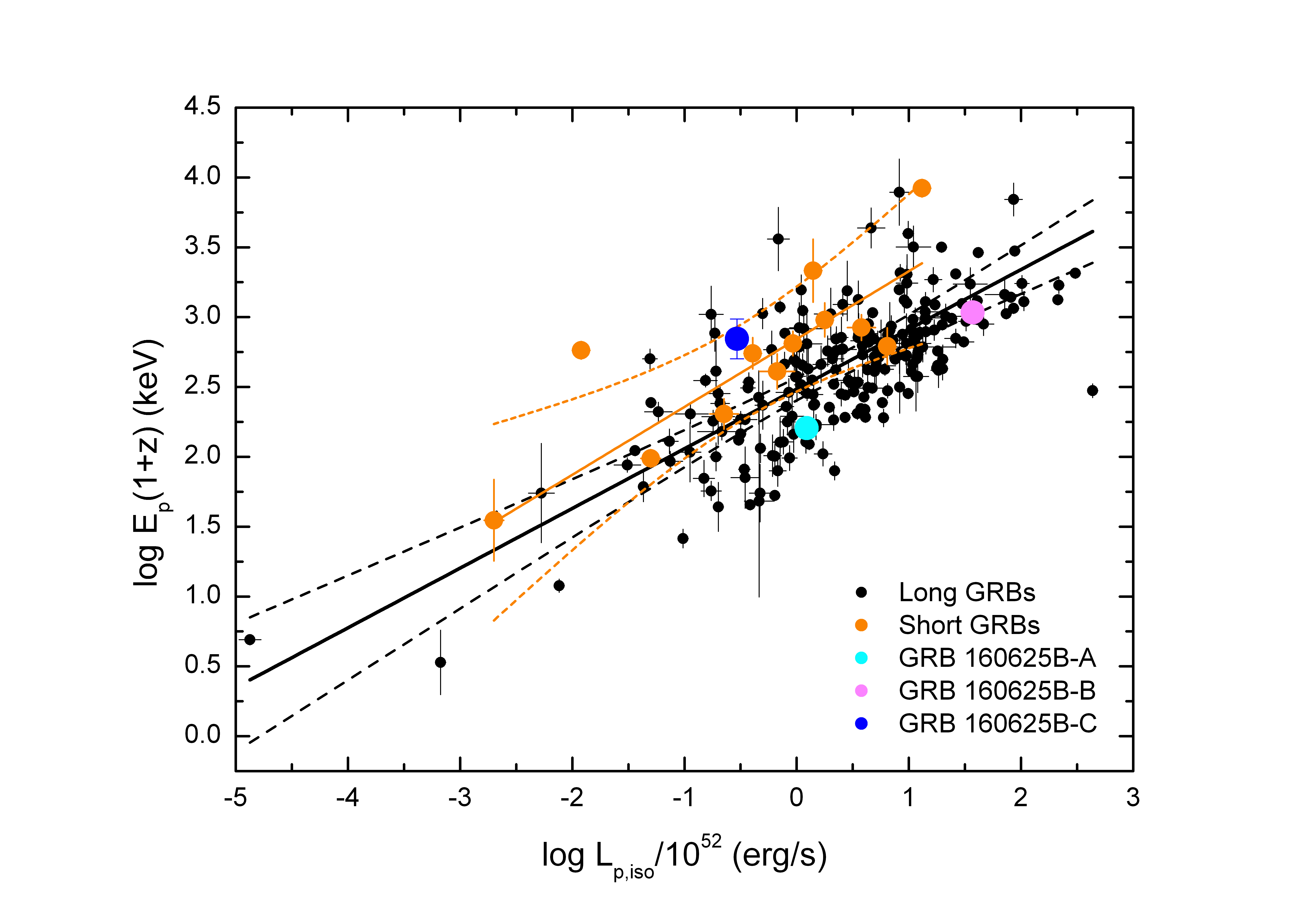}\\
\end{tabular}
\caption{{\bf The three sub-bursts in GRB 160625B GRB correlaiton diagrams.} All three sub-bursts are consistently located in the typical long (or Type II) GRB regions.  {\bf a:} GRB 160625B in the $E_p$-$E_{\rm iso}$ diagram.The solid lines are the best fit correlations: log$E_{\mathrm{p}}/(1+z)=(3.24\pm0.07)+(0.54\pm0.04)$log$(E_{\mathrm{iso}}/10^{52})$ for short GRBs, log$E_{\mathrm{p}}/(1+z)=(2.22\pm0.03)+(0.47\pm0.03)$log$(E_{\mathrm{iso}}/10^{52})$ for long GRBs. {\bf b:} GRB 160625B in the $E_p$-$L_{\rm iso}$ diagrams. The solid lines are the best fit correlations: log$E_{\mathrm{p}}/(1+z)=(2.84\pm0.09)+(0.49\pm0.07)$log$(L_{\mathrm{iso}}/10^{52})$ for short GRBs and log$E_{\mathrm{p}/(1+z)}=(2.49\pm0.03)+(0.43\pm0.03)$log$(L_{\mathrm{iso}}/10^{52})$ for long GRBs.All error bars represent 1-$\sigma$ uncertainties.
}
\label{fig:epeisoepliso}
\end{figure*}

One distinct feature of GRB 160625B is that it is composed of three sub-bursts separated by two quiescent times ($\sim 180$ s and $\sim 339$ s, respectively; Figure 1), {which, in combination with the different spectral properties of those sub-bursts, provides us an unique opportunity to study the transition of jet composition (see Method for details)}. The first sub-burst, which is undoubtedly related to the main event (the second sub-burst; see Supplementary Figure 1), is particularly interesting as its short duration ($\sim 0.84$ s) gives it the appearance of a traditional short GRB (Figure 1). To test such a possibility, we over-plot all the three sub-bursts on the $E_p-E_{\rm iso}$ and $E_p-L_{\rm p, iso}$ diagrams in Figure \ref{fig:epeisoepliso}, where $E_p$ is the peak energy in the $\nu F_\nu$ energy spectrum, $E_{\rm iso}$ is isotropic equivalent radiated energy and $L_{\rm p, iso}$ is the isotropic peak luminosity. The values of $E_p$ in Figure \ref{fig:epeisoepliso} are from the time-integrated spectral fitting using a Band function or cutoff power law model, which are listed in Table 1. It can be seen that all three sub-bursts lie in the track of long GRBs in both diagrams. According to the multi-wavelength physical classification criteria flowchart (Figure 8 of \cite{Zhang2009}), all three sub-bursts fall into the correlations of Type II (massive star origin) rather than Type I (compact star origin) GRB category. On the other hand, a good fraction of long GRBs show a short precursor before the main emission episode \cite{Hu2014}. The first sub-burst in our analysis is consistent with being a precursor of the main burst.

{ We study the time-resolved spectra of the first sub-burst by dividing the sub-burst into 6 slices with a bin size of 0.22 s. {Such a bin size ensures that there are at least 20 net counts in at least 10 energy channels in the spectrum.} We first fit the time-resolved spectra in each slice with a power-law plus an exponential cutoff (hereafter CPL) model and found that the lower energy photon index is close to 0 (-0.16 to 0.24) (Supplementary Table I). Such a photon index is much harder than the typical low-energy photon index ($\sim$ -1) of long GRBs, and is beyond the so-called synchrotron line of death ($\alpha = -2/3$) \cite{Preece2000}. Such a spectrum suggests a significant contribution of thermal emission from the fireball photosphere \cite{Meszaros00,Ryde2010}. Physically, the photosphere emission should be a multi-color blackbody, since at an observing time, the observer receives emission from different locations in the equal arrival time volume \cite{Peer08,DengZhang14}.
We are then motivated to fit the observed spectrum with a multi-color blackbody model.
As shown in Supplementary Table I, such a model indeed gives the best fit in comparison with the power law and cutoff-PL models based on the Bayesian Information Criteria (BIC). We thus suggest that the first sub-burst exhibits thermal spectra. 
The single blackbody model can also reasonably fit the data. We also apply this model to derive the characteristic temperature of the photosphere emission, which is used to derive physical parameters of the photosphere.}
While a few GRBs have previously been found to have a thermal precursor \cite{ 1991Natur.350..592M}, GRB 160625B allows one for the first time to perform a fine time-resolved spectral analysis and track the blackbody evolution of a precursor. Based on the standard fireball photosphere model \cite{Peer2007}, one can infer the fireball launch radius $r_0 Y^{3/2}$ and the bulk Lorentz factor $\Gamma Y^{-1/4}$, where $Y>1$ is the ratio between total energy of the fireball and the energy emitted in $\gamma$-rays at the photosphere. The derived values are presented in {Supplementary Table II}, which suggests a highly relativistic fireball launched from a radius of a few $\times 10^8$ cm, falling in the range of a jet emerging from a Wolf-Rayet progenitor star \cite{Alberto2010}.

The 2nd sub-burst is consistent with a typical long GRB with $T_{90}\sim 35.1$ s. At redshift $z=1.406 \pm 0.001$ determined by our Gran Telescopio CANARIAS (GTC) observations (see Supplementary Figure 2), the derived total isotropic $\gamma$-ray energy is $E_{\rm iso}\sim 3\times 10^{54}$ erg, making it the fifth most energetic GRB ever observed with known redshifts. Thanks to its high flux, one can divide the 2nd sub-burst into { 71} slices { based on the criterion that each spectrum {on average has 20 net counts in each of the 128 energy channels (so a total of $\sim$ 2500 net counts)} to conduct a fine time-resolved
spectral analysis}. The lengths of the time bins are comparable to those used for the 1st sub-burst. The spectrum of each time slice cannot be fit with a Planck-like spectrum, but can be successfully fit by the analytic Band function model\cite{Band1993}. The best-fit parameters are listed in Supplementary Table 1 and their evolution with time is plotted in Figure 1. Adding an additional thermal component neither improves the goodness of the fit, nor leads to constraints on the parameters of the new component (for details see Supplementary Figure 4). The $E_p$ evolution first shows a hard-to-soft evolution trend during the time 180-190 s interval following the first trigger, and then generally follows an intensity tracking pattern thereafter. All the Band-function parameters are typical and similar to most long GRBs. { Physically, the above typical GRB spectrum can be explained by synchrotron emission in a decaying magnetic field \cite{Uhm2014}, and confronting the model to the data indeed suggests that such a model can fit the data equally well as the Band function \cite{Zhang2016}. We are then motivated to apply the synchrotron model fit as described in \cite{Zhang2016} to fit the time-dependent spectra of the second sub-burst. For each time slice, we fit the data with both the Band model and the physical synchrotron model. As shown in Supplementary Table 1, the synchrotron model fits give comparable statistics values (PGSTAT) as the Band function model, suggesting that the emission of the second sub-burst is synchrotron-dominated.

One example of such synchrotron fits is presented in Supplementary Figure 5. The non-detection of a thermal component in these time bins suggests a Poynting flux dominated outflow with suppressed photosphere emission. }

Although six times longer than the 2nd sub-burst, the 3rd sub-burst is the faintest among the three sub-bursts -- about 126 times weaker than the 2nd sub-burst in terms of peak flux (Table I). Due to low photon statistics, most time resolved spectra of the 3rd sub-bursts can be adequately fit by a simple power law model with the exception of two slices, in which the CPL model is needed and $E_p$ can be constrained (Supplementary Table I). The spectral evolution of the 3rd sub-burst is also shown in Figure 1. 

LAT detected high-energy emission up to 15.3 GeV during the prompt emission of GRB 160625B. Supplementary Figure 6 displays the energy-dependent light curves of the three sub-bursts with the $>1$ GeV LAT photons plotted individually as a function of arrival time. 

Optical observations (including the data presented here) using a dozen telescopes (see Supplementary Table II) were also carried out before the trigger of the first sub-burst, during the prompt and afterglow emission phases, which place deep upper limits during the quiescent phase between the first two sub-bursts and lead to detections during the 2nd sub-burst. There is a $\sim 3$ s lag between optical emission and the GBM band emission (Supplementary Figure 7). The linear polarization discovered in the optical emission of this burst \cite{Troja2017} is consistent with our conclusion that the outflow is Poynting-flux-dominated during the second sub-burst.

The spectral energy distributions (SEDs) from optical to GeV band at three epochs during the 2nd-sub-burst are presented in Supplementary Figure 8. An immediate impression is that the optical emission is in excess of the extrapolation of sub-MeV spectral component to the optical band. A similar case was observed in the ``naked-eye" GRB 080319B \cite{Racusin08}. A possible explanation could be that the optical emission is due to the synchrotron radiation while the sub-MeV emission is due to the synchrotron-self-Compton (SSC) emission \cite{Racusin08}.
However, such a model would predict a strong second-order SSC appearing in the GeV-TeV range \cite{Piran09}. {The LAT emission is below the extrapolation of the sub-MeV emission to the GeV range, without evidence of the second SSC component. This disfavors the SSC origin of the sub-MeV component. Instead we interpret the sub-MeV emission as synchrotron emission in a magnetic dissipation region within a Poynting-flux-dominated outflow \cite{ZhangYan11, Uhm2014}.} The optical emission would be produced at a farther distance from the central engine, as suggested by the $\sim 3$s time delay. One possibility is the reverse shock emission as the variable moderately magnetized ejecta (after significant magnetic dissipation to produce the sub-MeV emission itself) catch up in the wake of the initial fireball blastwave. 

{The three dramatically different isolated episodes with dramatically different time-resolved spectral features make the extraordinarily bright GRB 160625B an unique event to study significant changes in the jet composition among different emission episodes in a single GRB}. 
So far the cases of a fireball characterized by a multi-temperature blackbody or even a Planck function in the time-resolved spectra (e.g, GRB 090902B) \cite{Ryde2010}, a Poynting flux dominated jet characterized by a Band-only spectrum (e.g. GRB 130427A and GRB 130606B) \cite{Preece2014, Zhang2016}, and a hybrid jet characterized by the superposition of a sub-dominant thermal component and a dominant synchrotron component have also been observed (e.g. GRB 100724B, GRB 110721A, and GRB 120323A) \cite{Axelsson2012}.
Some bright bursts (e.g, GRB 911118 \cite{Ghirlanda2003}) exhibit thermal-like time-resolved spectra at the beginning or around the peak of the light curves followed by non-thermal time-resolved spectra within a single emission episode, which may indicate a gradual change of the jet composition. {GRB 090902B showed a widening of the spectrum with time \cite{Ryde2010} which may be understood as a gradual change of jet composition. However, it is only GRB 160625B that shows such a dramatical change of jet composition in such widely separated emission episodes.}

The apparent change of jet composition between two emission episodes of GRB 160625B sheds light on the central engine of the GRB. The first sub-burst is likely associated with the initial iron core collapse that forms a black hole or a rapidly spinning magnetar surrounded by some high-angular momentum material. The initial hyper-accretion onto the central object forms a matter-dominated fireball via neutrino-anti-neutrino ($\nu\bar\nu$) annihilations. The jet may be intermittent, but lasts longer than the duration for the jet to penetrate through the star. The first several pulses within the ejecta merge and eventually break out of the star to make a successful thermally-dominant GRB precursor \cite{Wang07}. After the early prompt accretion phase, the central engine enters a quiescent phase lasting for $\sim 180$ s. This could be due to one of the following reasons: 1). In the context of the black hole central engine model, the accretion may be halted by a magnetic barrier from the central engine \cite{Proga06} after the initial prompt accretion phase. One needs to wait to accumulate enough materials to break through the barrier. The free-fall time scale of the C/O core is $\sim 100$ s, which may be related to the waiting time before the magnetic barrier is broken \cite{Wang07}. 2). In the context of the magnetar central engine model, after the initial accretion phase, the proto-magnetar would take some time to cool before a Poynting-flux-dominated outflow is launched \cite{Metzger11,2007MNRAS.382.1029K}. The cooling time scale is typically 10s of seconds, somewhat shorter than the quiescent period. However, the competition between fall-back spin-up and magnetic spin-down may delay the launch of the outflow. In both the above-mentioned scenarios, the central engine is strongly magnetized when the second jet is launched, so that the emission would be characterized by a synchrotron spectrum without a significant thermal component from the photosphere. After the main emission episode (the second sub-burst), the central engine would reactivate again similar to the GRBs that show significant late-time X-ray flares \cite{Burrows05}, which powers the third emission episode (the third sub-burst). The Blandford-Znajek mechanism is insensitive to the accretion rate and can be responsible for the late activity of the central engine which would explain the main phase and the extended emission \cite{2007MNRAS.382.1029K}.
Unfortunately due to observational constraints Swift XRT observations were not available until 10$^4$ s after the trigger, so we could not test the X-ray flare hypothesis. Similar behavior was also seen in GRB 130427A, which showed a weaker extended emission accompanied by a bright X-ray flare after the main burst \cite{Ackermann14}.

Some GRBs have a precursor, while some others are characterized by extended emission. GRB 160625B consists of three distinct emission episodes which may correspond to the precursor, main burst, and delayed extended emission, respectively. Thanks to its exceptional brightness, one was able to study the detailed spectral evolution of the three episodes in detail. If the burst had been less energetic or had been placed at a larger distance, one would only detect two episodes or even only one, and the event would be ascribed as a ``normal'' GRB. It is possible that GRB 160625B is the proto-type of at least one sub-group of long GRBs, which display three distinct emission episodes \cite{Hu2014}, and the physical processes delineated in this paper may apply to these bursts in general.

\vspace{\baselineskip}
\textbf{References}

\makeatletter
\def\@biblabel#1{\@ifnotempty{#1}{#1.}}

\textbf{Acknowledgments}

BBZ thank Y.-Z. Fan, Y.-Z. Wang, H. Wang, K. D. Alexander and D. Lazzati for helpful discussions. We are grateful to K. Hurley, I. Mitrofanov, A. Sanin, M.
Litvak, and W. Boynton for the use of {\it Mars Odyssey} data in the triangulation. We acknowledge the use of the public data from the {\it Swift} and {\it Fermi} data
archives. BBZ and AJCT acknowledge support from the Spanish Ministry Projects AYA 2012-39727-C03-01 and AYA2015-71718-R. Part of this work made use of BBZ's personal IDL code library ZBBIDL and personal Python library ZBBPY. The computation resources used in this work are owned by Scientist Support LLC. BZ acknowledges NASA NNX14AF85G and NNX15AK85G for support. ZGD acknowledges the National Natural Science Foundation of China (grant No. 11573014). YDH acknowledges the support by China Scholarships Council(CSC) under the Grant No.201406660015. Mini-MegaTORTORA belongs to Kazan Federal University and the
work is performed according to the Russian Government Program of Competitive
Growth of Kazan Federal University. AP, EM, PM and AV are grateful to RFBR (grant
17-02-01388 ) for partial support. AP and SBP acknowledge joint BRICS grant RFBR 17-52-80139 and 388-ProFChEAP for partial support. RI is grateful to the grant RUSTAVELI
FR/379/6-300/14 for a partial support. Observations on Mini-MegaTORTORA are
supported by the Russian Science Foundation grant No. 14-50-00043. VAF and MAS are thankful to the Russian Science Foundation (grant 14-50-00043). LM and AFZ acknowledge support from INTA-CEDEA
ESAt personnel hosting the Pi of the Sky facility at the BOOTES-1 station. HG acknowledges the National Natural Science Foundation of China under grants 11603003. XYW acknowledges the National Natural Science Foundation of China
under grant 11625312. ZGD, XFW, BZ , XYW, LS and FWZ are also supported by the 973 program under
grant 2014CB845800. FWZ is also supported in part by NSFC under grants No. U1331101 and No. 11163003 and by the Guangxi Natural Science Foundation (No. 2013GXNSFAA019002) and by the project of outstanding young teachers' training in higher education institutions of Guangxi. LS acknowledges the support by the National Natural Science Foundation of China (No. 11103083) and the Joint NSFC-ISF Research Program (No. 11361140349). SRO acknowledges the support of the Leverhulme Trust. SJ acknowledges the support from Korea Basic Science Research Program through NRF-2014R1A6A3A03057484 and NRF-2015R1D1A4A01020961, and IHP from National Research Foundation grants of No. 2015R1A2A1A01006870 and No. 2015R1A2A1A15055344. RA, DF, and DS gratefully acknowledge support from RSF grant 17-12-01378. AK acknowledge the Science and Education Ministry of Kazakhstan (grant No. 0075/GF4).

\vspace{\baselineskip}

\vspace{\baselineskip}
\cleardoublepage

\noindent \textbf{\large METHODS}

\vspace{\baselineskip}

\noindent\textbf{Uniqueness of GRB 160625B} GRB 160625B is unique in terms of the following two aspects: (1) it has three well-separated emission regions (the first, second and third sub-bursts);
(2) The first and second sub-bursts have distinct spectral types (thermal and non-thermal). These two properties together make GRB 160625B an unique event and provide 
a solid identification of the transition of jet composition thanks to the high brightness and long gap between the sub-bursts. When comparing with some GRBs in previous studies, we notice that 
there are some GRBs showing well-separated emission episodes ( e.g, GRB 070721B \cite{2007GCNR...73....2Z}, GRB 091024 \cite{2011A&A...528A..15G}, GRB 110709B \cite{2012ApJ...748..132Z}) but none of those bursts
showed the thermal-to-non-thermal transition before and after their gaps. We also noticed that there were some GRBs which show thermal to non-thermal transition (e.g, GRB 911118 \cite{Ghirlanda2003}, GRB 090902B \cite{2009ApJ...706L.138A,Ryde2010}), but such transition is continuous within one or two overlapping emission regions without any quiescent gaps. Moreover, the first sub-burst of GRB 160625B can be regarded as a precursor. Theoretically a thermal precursor leading a main non-thermal jet emission component has been 
expected \cite{Wang07,Burlon2008}, but has not
been identified in the observations \cite{Nora2010,Lazzati05}.
GRB 160625B is such a unique event with a very bright precursor which allowed 
us to identify its thermal emission in a time-resolved manner.

\noindent\textbf{Data Reduction and Spectral Fitting} We processed the GBM and LAT data using the standard software as described in \cite{Zhang2011} and \cite{Zhang2016}. The data from the two brightest Sodium Iodide (NaI) scintillation detectors and the brightest Bismuth Germanate (BGO) scintillation detectors on board {\it Fermi} are used for our spectral fitting analysis. The task of spectral fitting is performed using our software package, {\it McSpecFit}, which {combines} a Bayesian MC engine {\it McFit}, general forward-folding codes, and likelihood calculations. The package can directly test any model against the observed spectra. In particular, the Bayesian Monte-Carlo engine ({\it McFit}), as described in \citep{zhangbb15_zbb, Zhang2016}, employs a Bayesian Monte-Carlo (MC) fitting technique to reliably fit parameters that are constrained by the data even {when} other parameters are unconstrained. Using this technique, the best-fit parameters and their uncertainties can be realistically determined by the converged MC chains. The general forward-folding codes are used to compare a theoretical spectral model to data. In general, one needs to convolve the model with instrumental response, the Detector Response Matrix (DRM), to compare the models with the observational data. A general-purpose forward-folding code has been written\cite{Zhang2016} and it can handle the instrument response $R(I, E)$ from any mission (e.g. {\it Fermi}/GBM), read in any model spectrum, $F(E, P)$, regardless of its form, and calculate the model-predicted count spectrum $C_M(I, P)=\int_{0}^{\infty}F(E, P)R(I, E)dE$. Such $C_{\rm M(I, P)}$ can be directly compared with the observed count spectrum $C(I)$. We then calculate the likelihood for those $C_M(I, P)$ and $C(I)$ pairs. In this paper, we use the maximum likelihood-based statistics, the so-called PGSTAT, for Poisson data, given by \cite{cash79}, but assuming the background to be Gaussian (see also \url{https://heasarc.gsfc.nasa.gov/xanadu/xspec/}). Such a likelihood function is then input into our fitting engine {\it McFit} so that the parameter space can be explored. { The priors of the fitting free parameters are set to uniform (those listed as {\it``parameter name"} in Supplementary Table I ) or log uniform (those listed as {\it``log parameter name"} in Supplementary Table I) distributions in physically-allowed large ranges{~(see below)}}. 

{ LAT data are used and presented in the temporal analysis (see Figure 2 
and Supplementary Figure 6). LAT data are also presented in the spectral
energy distribution (SED) from optical to LAT energies in each optical
observational bins, where we found that the LAT emission is below the
extrapolation of the sub-MeV emission to the GeV range, without the evidence
of the second SSC component. This disfavors the SSC origin of the
sub-MeV component.
LAT data are not used in our fine-time resolution spectral analysis.
This is because (1) the inference of jet composition mostly depends on the
low-energy photon index in the sub-MeV range; (2) LAT data are only available
in some sub-bursts but not all sub-bursts; and (3) for our fine time slices
(as small as 0.05 s), the number of LAT-band photons is usually zero or very small number close to one,
including the LAT data would not alter our spectral fitting results in MeV range nor significantly constrain the parameters in the GeV range .}

\medskip

\noindent\textbf{Synchroton Model} We employ the fast-cooling synchrotron emission model with a decaying magnetic field\citep{Uhm2014} to fit the time-resolved spectra. The flux density (in unit of mJy) of such model can be calculated numerically in the form
$$ F_\nu(E)=M(E; \Gamma, p, \gamma_{\rm inj},R_{\rm inj}^0 , q , B_0 ,b ,\hat t, z),$$
which includes the following parameters:
\begin{itemize}
\item the bulk Lorentz factor $\Gamma$ of the emitting region;
\item the magnetic field strength $B_0$ {(in units of {\it Gauss})} at $10^{15}$ cm;
\item the magnetic field decay index $b$;
\item the power-law index $p$ of the injected electron spectrum;
\item the minimum injection Lorentz factor $\gamma_{\rm inj}$ of electrons;
\item the power-law index $q$ of the injection rate as a function of time;
\item the {normalization} of the electron injection rate $R_{\rm inj}^0$ {in units of s$^{-1}$} ;
\item the observer-frame time since the first electron injection $\hat t$ {in units of s} of a given emission episode;
\item the redshift $z$.
\end{itemize}.

The details of model calculations can be found in \cite{Uhm2014}. The redshift $z$ is fixed to 1.406 for GRB 160625B. Other parameters are allowed in the following ranges - 
$\Gamma$ : [$10^2$, $10^3$]; $p$ : [2, 6]; $\gamma_{\rm inj}$: [$10^{3}$, $10^{6}$];
$R_{\rm inj}^0$ ($s^{-1}$): [10$^{44}$, 10$^{50}$];
$q$: [0.01, 5];
$B_0$ (G): [5, 1000];
$b$: [0.9, 2.0] ;
$\hat t$ (s): [0.1, 10.0].

\noindent\textbf{Method to Determine the Lag between Optical and $\gamma$-ray Bands} 
The time bins in optical data are
10 s (Pi of the Sky) and 30 s (Mini-MegaTORTORA). There are only 4-5
bins in the optical band. The bin size of the gamma-ray light curve is much
smaller ($\sim$ 50 ms). The traditional Cross-Correlation Function (CCF)
method is not applicable here to determine the lags due to the smaller number of
bins and different bin sizes in two energy bands. Instead, we determined the
lag by the following procedure: (1) shift the gamma-ray light curve by T$\rightarrow$T+$\Delta$T;
(2) re-bin the shifted gamma-ray light curve to the exact optical bin size;
(3) calculate the ``similarity" between the binned-and-shifted gamma-ray 
light curve and the optical light curve. We use a $\chi^2$ value defined as
$\chi^2=\Sigma\frac{(F_{opt}-F_\gamma)^2}{\sigma_{opt}^2+\sigma_\gamma^2
}$ to define the goodness of the match; (4) repeat (2) and (3) by varying $\Delta$T from -30 to 30 s;
(5) determine the lag by searching which $\Delta$T gives a minimal $\chi^2$.
The derived 1-$\sigma$ uncertainties of the lag is estimated by comparing 
$\chi^2$ values to the minimum one using F-test. We note that the similar approach was used by \cite{Tang2006} for GRBs 990123 and 0401219.

A similar $\sim$ 2 s delay was also observed in the ``naked-eye" GRB~080319B \cite{Beskin2010}. The optical emission of GRB 160625B may have originated from reverse shocks, residual internal shocks \cite{LiWaxman08} or from free neutron decay at a large radius from the central engine \cite{Fan09}.

\makeatletter
\apptocmd{\thebibliography}{\global\c@NAT@ctr 30\relax}{}{}

\makeatother

\begin{table*}
\begin{scriptsize}
\caption{Properties of Three Sub-bursts in GRB 160625B.}
\label{tab:sum}
\begin{tabular}{|c|c|c|c|}
\hline
Sub-burst & A & B & C \\
\hline

Duration: T$_{90}$ [s] (15 - 350 keV )& $0.84^{+0.03}_{-0.01}$&$35.10^{+0.13}_{-0.23}$ &$212.22^{+0.95}_{-2.27}$ \\
waiting time [s] & - & $\sim$ 180 & $\sim$ 339 \\
Time lag $(15-25:50 -150 keV)$ & 0.007 $\pm$ 0.013 & $0.80 \pm 0.05$ & -$0.06\pm 0.30$ \\
Time-integrated peak energy E$_p$ (keV) & 66.8$\pm$ 1.8 & $448.8_{-4.22}^{+4.42} $ & $290.5_{-55.7}^{+95.0} $ \\
Energy fluence $f$ (erg cm$^{-2}$) & $1.75\pm 0.05 \times 10^{-6}$ & $6.01\pm 0.02 \times 10^{-4}$ & $5.65\pm 0.02 \times 10^{-5}$ \\
Peak energy flux F$_p$ (erg cm$^{-2} $ s$^{-1}$) & $2.42\pm 0.11 \times 10^{-6}$ & $7.31\pm 0.13 \times 10^{-5}$ & $5.80\pm 0.13 \times 10^{-7}$ \\
Isotropic Energy E$_{\rm iso}$ (erg) & $8.86\pm 0.24 \times 10^{51}$ & $3.05\pm 0.01 \times 10^{54}$ & $2.87\pm 0.01 \times 10^{53}$ \\
Isotropic peak luminosity L$_{\rm p, iso}$ (erg s$^{-1}$) & $1.23\pm 0.05 \times 10^{52}$ & $3.71\pm 0.06 \times 10^{53}$ & $2.94\pm 0.07 \times 10^{51}$ \\

\hline

Redshift $z$ & \multicolumn{2}{c}{1.406$\pm$0.001} & \\
\hline
Most energetic LAT photons & \multicolumn{2}{c}{15.3 GeV (346.2 s); 6.95 GeV (793 s); 5.4 GeV (573 s)} & \\

\hline

\hline
 
\end{tabular}
\end{scriptsize}

\end{table*}

\newpage
\clearpage
\beginsupplement

\vspace{\baselineskip}

\vspace{\baselineskip}
\textbf{Supplementary Information}

\textbf{1. Additional gamma-ray observations of GRB 160625B and localization by the IPN}.

GRB 160625B was detected by four InterPlanetary Network (IPN) experiments -- {\it Fermi}-GBM in a low-earth orbit, {\it INTEGRAL} SPI-ACS\cite{S18}, in a highly elliptical orbit, Konus-{\t Wind}\cite{S19}, in orbit around the Lagrangian point L1, and {\it Mars-Odyssey} HEND\cite{S20}, in orbit around Mars -- at 0.02, 0.38, 4.32, and 276.46 light-seconds from Earth, respectively. The burst showed three emission episodes: the initial short duration episode (observed by {\it Konus}, {\it Fermi}-GBM, SPI-ACS), the main very intense episode (observed by all IPN instruments and Fermi-LAT), and the final rather weak episode (observed by GBM, SPI-ACS and Konus).

As the two first sub-bursts are entirely different, we cannot exclude that they could be related to two independent events whose wavefronts happened to arrive at Earth 180 s apart. In order to check this we have derived the annuli for the first two sub-bursts (see Supplementary Figure \ref{fig:ipn}) and found that both of them are consistent with the Swift-XRT position for GRB 160625B, hence confirming that they very likely have the same origin.

As observed by {\it Konus}-Wind, the second sub-burst had a fluence of (9.50$\pm 0.15$)$\times$10$^{-4}$ erg cm$^{-2}$, and a 256-ms peak flux, measured from T0+188.928 s, of 1.27$\pm$ 0.07$\times$10$^{-4}$ erg cm$^{-2}$s$^{-1}$ (both in the 20 keV - 10 MeV energy range), which are consistent with {\it Fermi} observations as listed in Table I.

\textbf{2. Simultaneous Optical Observations of GRB 160625B}.

\textbf{2.1. Very-wide field observation by CASANDRA cameras on the BOOTES Network}

Early upper limits were obtained by the CASANDRA all-sky cameras\cite{2008SPIE.7019E..1VC} at the BOOTES-1 and BOOTES-2 astronomical stations\cite{1999A&AS..138..583C} which were gathering data (as usual) with a 60s time resolution. As the GRB position on the sky was not at optimal conditions, only the upper limits could be derived, although the afterglow is marginally detected at maximum brightness.

\textbf{2.2. Mini-MegaTORTORA Observation} 

Mini-MegaTORTORA nine-channel wide-field monitoring system\cite{S12} reacted\cite{S13} to precursor GBM event and started observing its error box 52 seconds after it and 136 seconds before LAT trigger. Due to the large size of the GBM error box, observations were performed in the ``widefield+deep'' regime, with channels simultaneously covering $\sim$30$\times$30 deg field of view with 30 s exposures in white light to achieve deepest detection limit. The system acquired 20 frames in such a regime, covering the time interval from $T_0$-136 to $T_0$+466 s, and it detected a bright optical flash on a frame coincident with LAT trigger time ($T_0$-15.9 - $T_0$+14.1 s), with a magnitude of about V=8.8 mag, which then brightened for about 0.1 mag. It then faded following nearly smooth power-law decay with slope of about - 1.6, down to V=12.2 in the last acquired frame. The images acquired prior to the LAT trigger do not display any object at that position down to about V=13.5 mag.
The system also observed the same location between $T_0$+1691 s and $T_0$+2264 s, acquiring 20 additional 30-s exposure frames. These frames do not display any transient at the position of GRB~160625B brighter than V=13.5 mag.

\textbf{2.3. Pi of the Sky Observation}

The Pi of the Sky is a system of wide field of view robotic telescopes,
which search for short timescale astrophysical phenomena, especially
for prompt optical GRB emission\cite{S14}. The system was designed
for autonomous operation, monitoring a large fraction of the sky to
a depth of $12^{m}-13^{m}$ and with time resolution of the order
of 10~seconds. Custom designed CCD cameras are equipped with Canon lenses
f=85~mm, f/d = 1.2 and cover $20^{\circ}$\rm{x}$20^{\circ}$ of the sky each. 
The prototype telescope with two cameras since 2004
(currently at San Pedro de Atacama Observatory) and the final system with
16 cameras on 4 equatorial mounts was completed in 2014 at the INTA
El Arenosillo Test Centre in Spain\cite{S15}.

Cameras of the Pi of the Sky observatory in Spain were not observing the position
of the GRB160625B prior to the first sub-burst. Observations started 
only after receiving Fermi GBM trigger 488587220, about 140 seconds prior to the second
sub-burst. As the position estimate taken from Fermi alert and used to position
the telescope was not very accurate, the actual position of the burst happened
to be in the overlap region of two cameras, CAM35 and CAM39, resulting in two
independent sets of measurements. Light curves from both cameras were reconstructed
using the LUIZA framework\cite{S16,S17}. Observations were performed in wide visible
band, with IR-cut and UV-cut filters only, and 10~s exposure time. Images were calibrated
to the reference stars from Tycho 2 using the transformation from Tycho to the
Pi of the Sky system given by: 
 \[ V_{PI} = V_T + 0.235 - 0.292 \cdot (B_T - V_T) \]
No object brighter than 12.4m (3 $\sigma$ limit) was observed prior to the second
sub-burst. An optical flash was identified on an image starting -5.9 s before the time
of the LAT trigger, brightening to about 8m on the next image and then becoming
gradually dimmer, fading below our sensitivity after about 400 s.

\textbf{3. Additional Optical Observations (photometry)}

The afterglow emission was observed
by a dozen of telescopes worldwide: the Zeiss-1000 (1.0m) TSHAO telescope in Kazakhstan, the AS-32 (0.7m)
AbAO telescope in Georgia, the AZT-33IK (1.5m) telescope in Mondy (Russia), the AZT-11 (1.25m) and ZTSh (2.6m) telescopes at Crimean Astrophysical Observatory (CrAO), the AZT-22 (1.5m) at Maidanak Observatory (Uzbekistan) of IKI/ISON GRB follow-up network \cite{S23}, the 6.0m BTA and Zeiss-1000 (1.0m) SAO telescopes at Zelenchuk (Russia) (see Supplementary Table \ref{tab:optobs} for the detailed optical observations carried out in this work). Photometric data was reduced and calibrated using IRAF package.

\textbf{4. Additional optical Observations (spectroscopy)}
The 10.4 m Gran Telescopio Canarias (GTC; Canary Island, Spain), equipped with the Optical System for Imaging and Low-Intermediate Resolution Imaging Spectroscopy (OSIRIS) instrument \cite{cepa2000} obtained optical spectroscopy at two epochs (2 and 14 days post-burst). Data was reduced and calibrated the usual way using IRAF and custom tools coded up in python. We clearly detect several absorption lines at both epochs, from which a redshift z = 1.406 $\pm$ 0.001 is derived, and an intervening absorption system at z =1.319 (see Supplementary Figure \ref{fig:gtcspec}).

\textbf{5. Additional Spectral Analysis Results}

Supplementary Figure ~\ref{fig:spec_exs} shows a selection typical spectral fits to each sub-burst. Supplementary Figure \ref{fig:addingbb} shows an example showing the effect of adding a blackbody component to the Band
function model when fitting to the data in the time bin between 187.23 s and 187.67 s. Supplementary Figure ~\ref{fig:syncfit} shows an example fit to the data in one time bin (between 187.23 s and 187.67 s) using the synchrotron model. Supplementary Figure ~\ref{fig:wave} shows the energy-dependent light curves of the three
sub-bursts. Supplementary Figure ~\ref{fig:opt_mev_lag} shows the correlations between GBM prompt $\gamma$-ray
(black; 15-350 keV ) and optical observations (red) made by Pi of the Sky
(Cam 35), Pi of the Sky (Cam 39) and Mini-Mega TORTORA. Supplementary Figure ~\ref{fig:sed} shows the spectral energy distribution (SED) from optical to LAT energies in
each optical observational bins during the second sub-burst. Supplementary Table~\ref{tab:specfitting} shows the time-resolved spectral fitting results. Supplementary Table~\ref{tab:optobs} shows the optical observations of GRB 160625B. Supplementary Table~\ref{tab:lag_opt_gbm} shows time lag of optical light curves with respect to
the $\gamma$-ray light curve in the 2nd sub-burst. Supplementary Table~\ref{tab:sed} shows Spectral fitting parameters for the three SED slices. 
\begin{figure*}

\includegraphics[keepaspectratio,clip,width=0.9\textwidth]{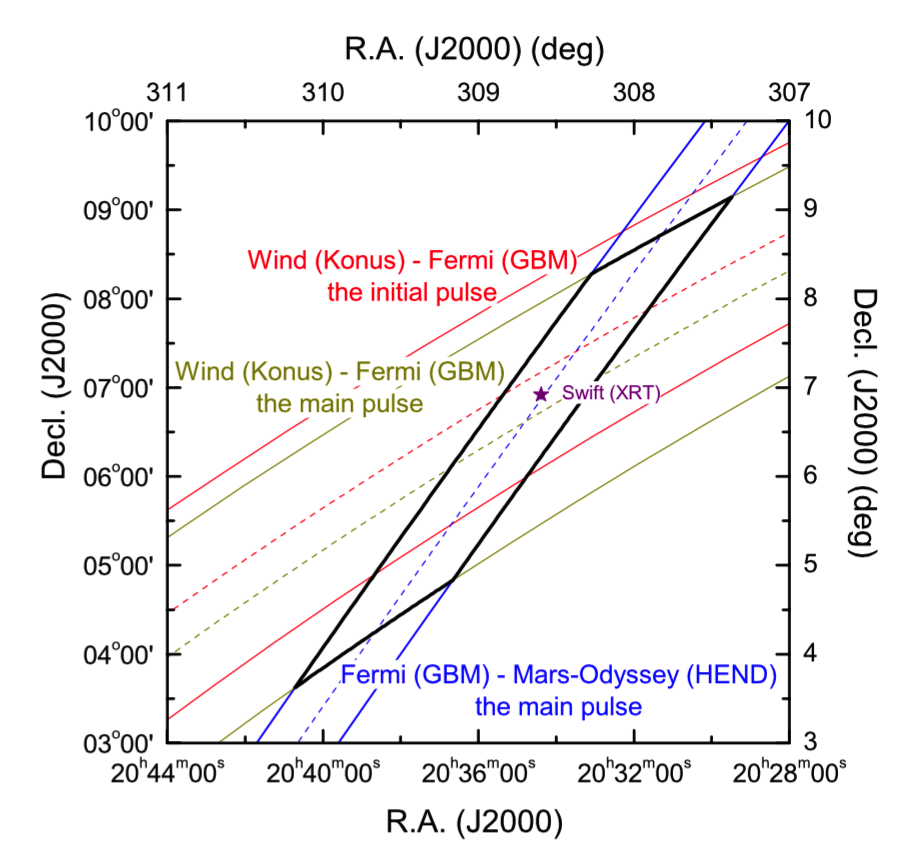}

\caption{The IPN triangulation of first two emission episodes of GRB 160625B. The 1.48 deg wide Wind-Fermi annulus derived for the first episode (1st sub-burst, 0.84 s duration) is consistent with {\it Swif}t-XRT position (indicated by the star) and the IPN error box defined by {\it Wind-Fermi} and {\it Fermi-Mars-Odyssey} annuli derived for the main episode (2nd sub-burst).}
\label{fig:ipn}
\end{figure*}

\begin{figure*}

\includegraphics[keepaspectratio,clip,width=0.6\textwidth]{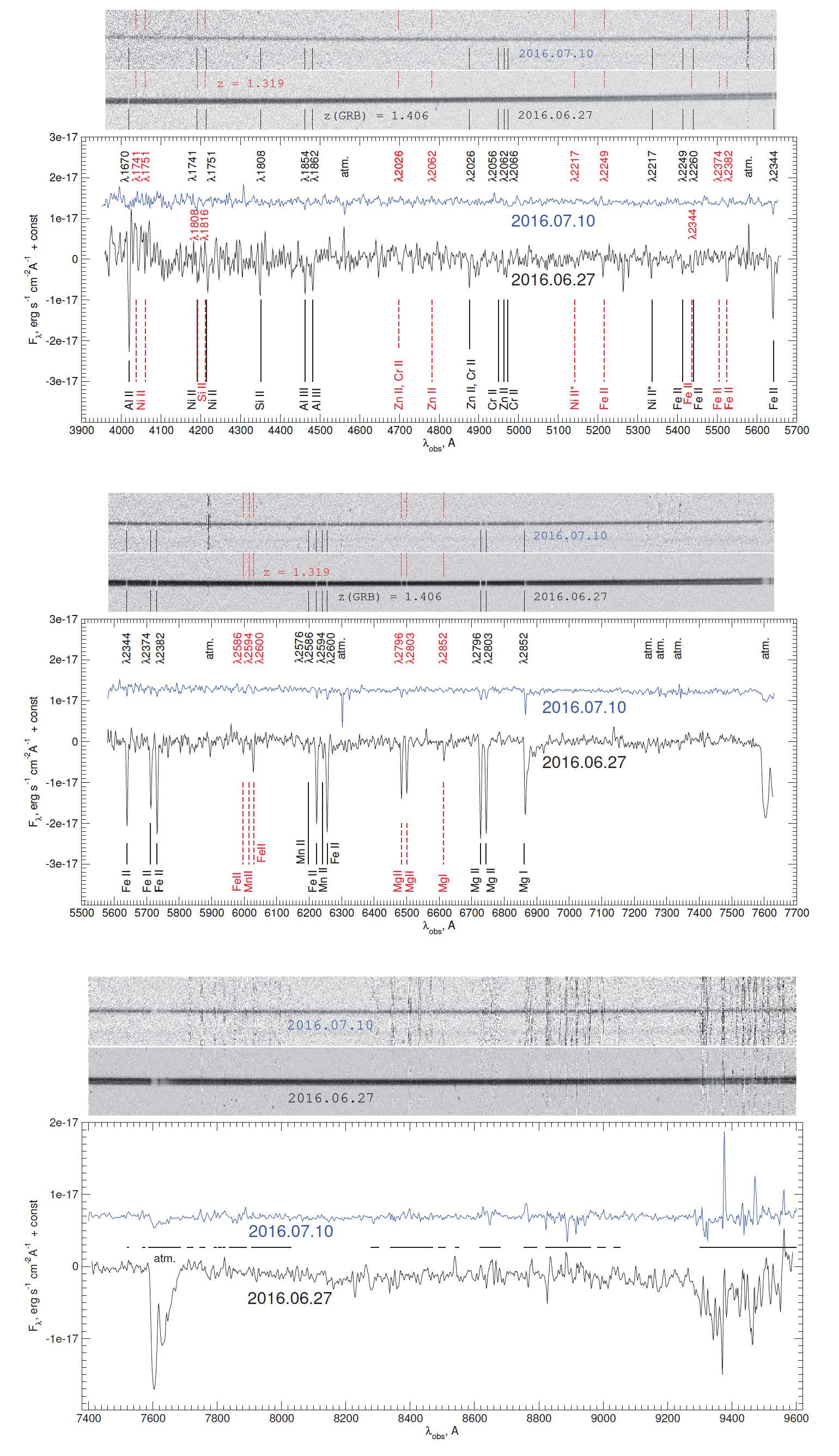}
\caption{ Spectra of GRB 160625B were obtained with the 10.4m GTC telescope (Osiris) on 2016.06.27 (bottom, black line) and
2016.07.10 (top, blue line). Absorption lines at z(GRB) = 1.406 are marked by solid black lines. An intervening
absorption system at z = 1.319 is marked by red dashed lines. Spectra was shoothed with the Gaussian core
(2.8 \AA, 4.4 \AA ~ and 3.2 \AA ~ for the blue, red and infrared parts respectively). Among the numerous atmospheric lines
in the infrared part of spectra (marked by the horizontal bands) we do not detect the [OII] lines at both
the GRB redshift or the intervening system (See also \cite{S3}).}
\label{fig:gtcspec}
\end{figure*}

\begin{figure*}

\begin{tabular}{ccc}
\includegraphics[keepaspectratio,clip,width=0.32\textwidth]{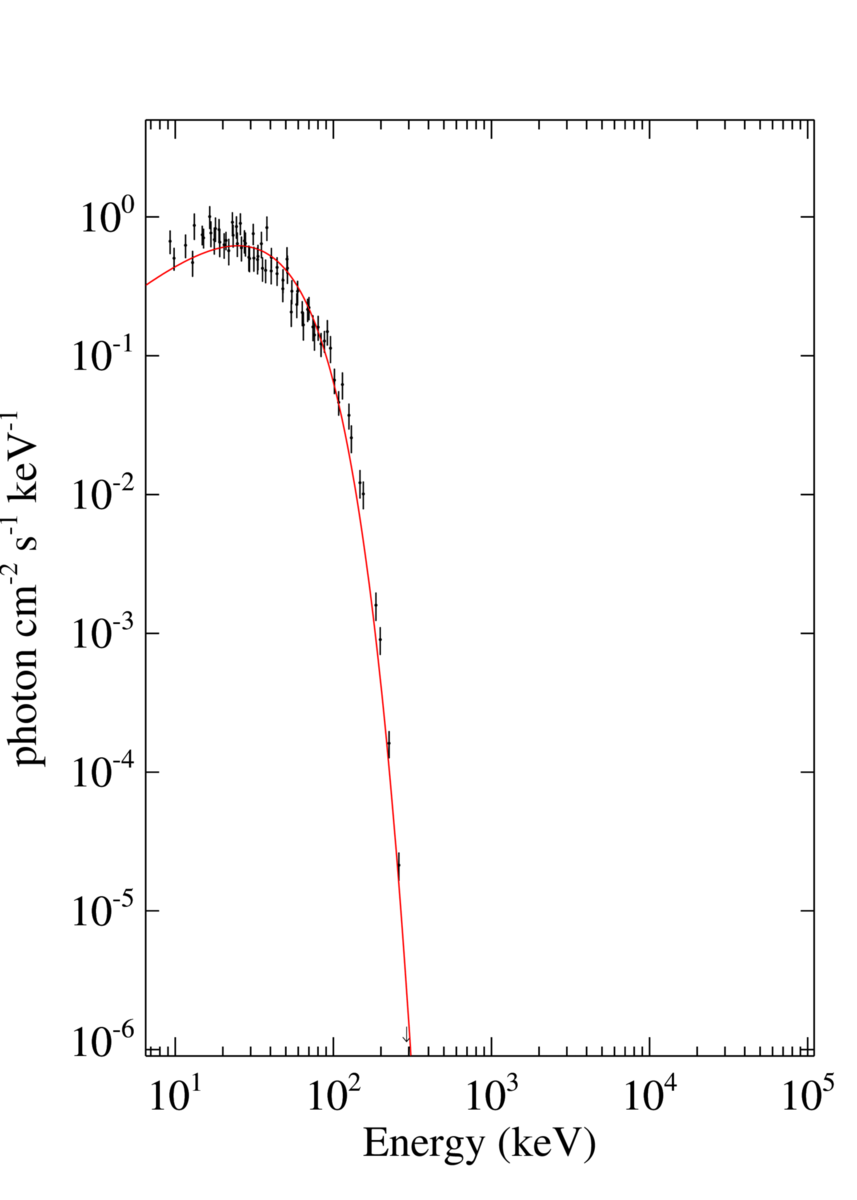} &\includegraphics[keepaspectratio,clip,width=0.32\textwidth]{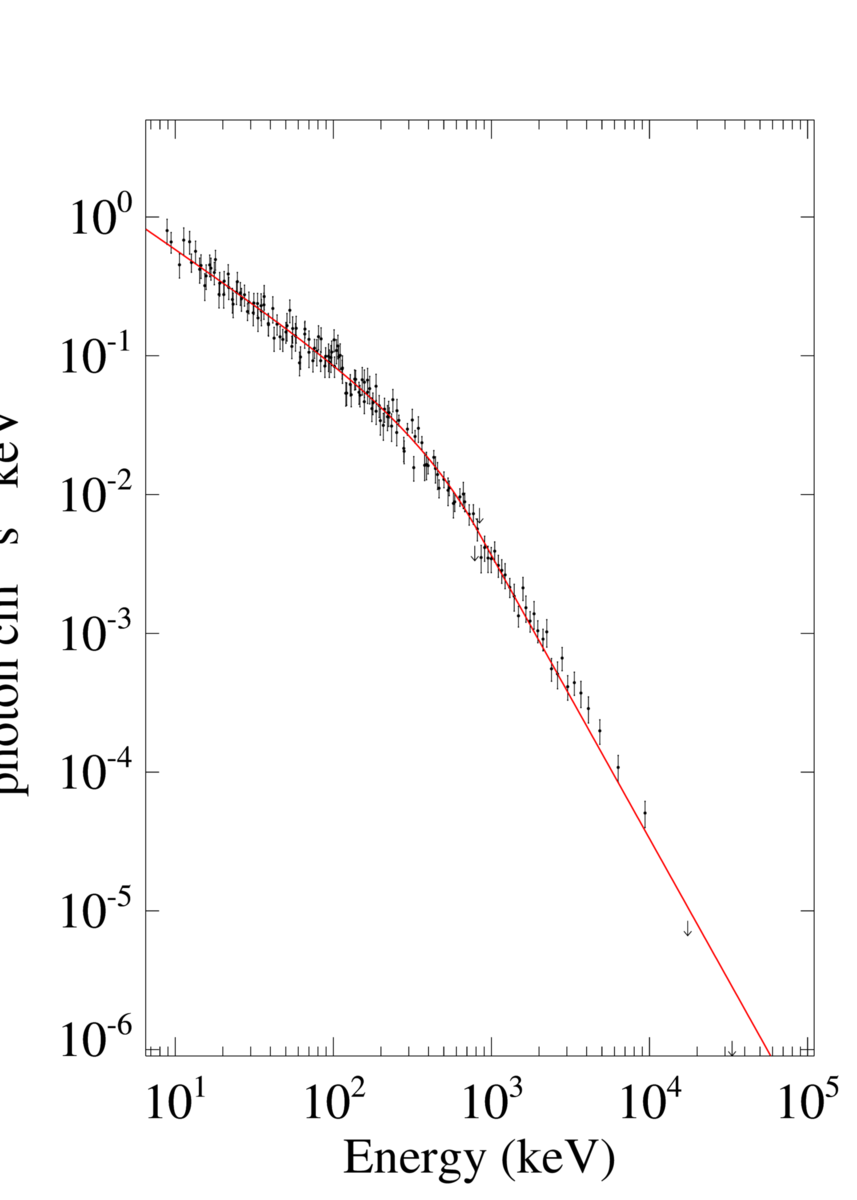}&\includegraphics[keepaspectratio,clip,width=0.32\textwidth]{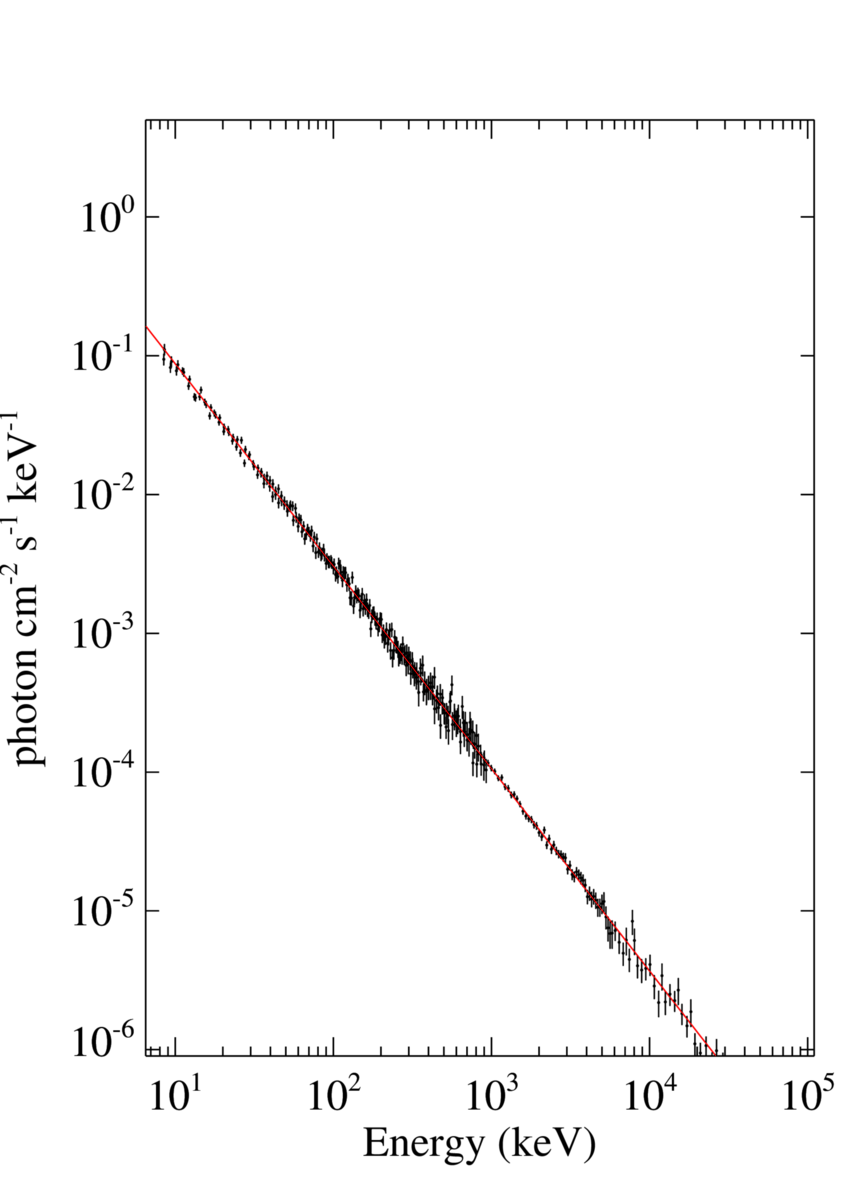} \\

\includegraphics[keepaspectratio,clip,width=0.32\textwidth]{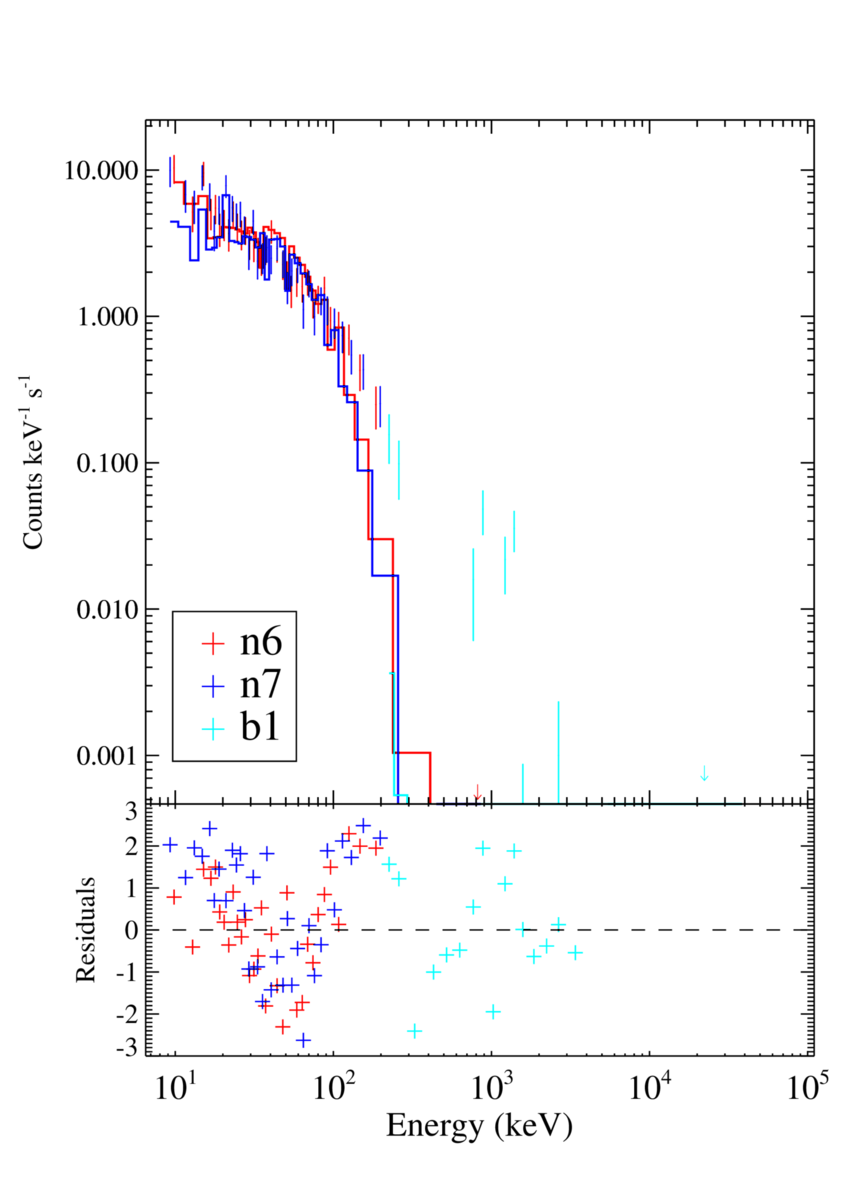} &\includegraphics[keepaspectratio,clip,width=0.32\textwidth]{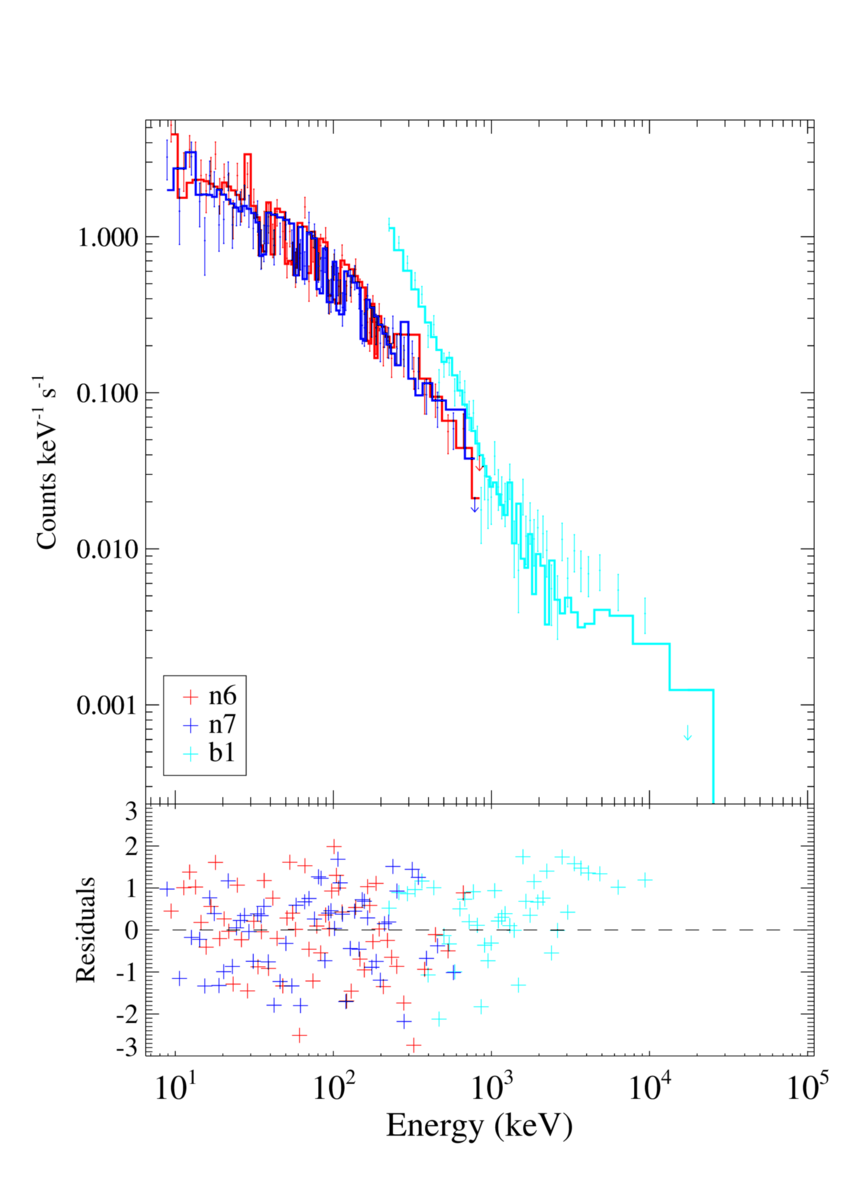}&\includegraphics[keepaspectratio,clip,width=0.32\textwidth]{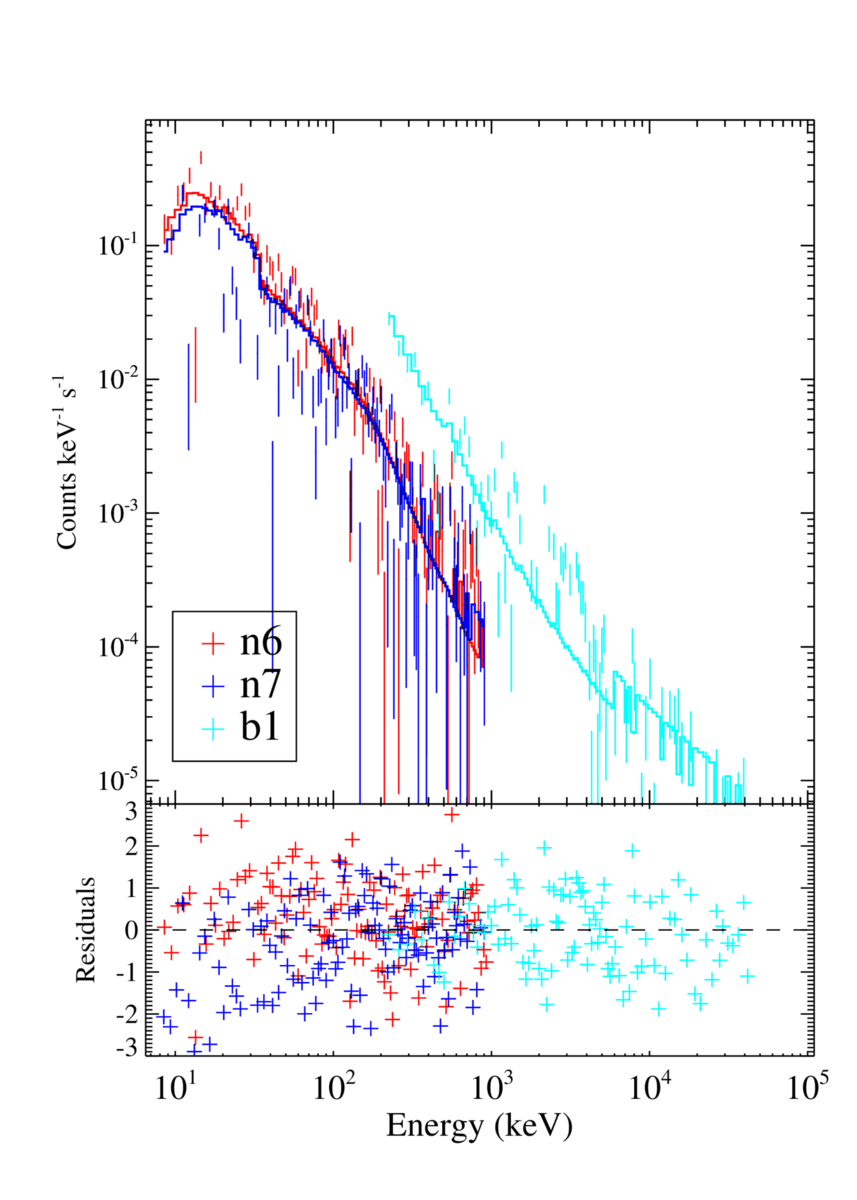} \\

\end{tabular}
\caption{ 
Typical spectral fits to each sub-burst. From left to right: black body fit of the 1st sub-burst between 0.13 and 0.35 s, Band function fit of the 2nd sub-burst between 187.23 and 187.67 s and power law fit of the 3rd sub-burst between 520.00 and 535.00 s. Top: photon spectra. Note the instrumental response has been deconvolved in each plot of the photon spectra. Bottom: Observed count rate spectra. The best-fit paramters can be found in Supplementary Table \ref{tab:specfitting}.
}
\label{fig:spec_exs}
\end{figure*}

\begin{figure*} \centering
\begin{tikzpicture}[
 every node/.style={anchor=south west,inner sep=0pt},
 x=1mm, y=1mm,
 ] 
 \node (fig1) at (0,0)
 {\includegraphics[width=5.1in]{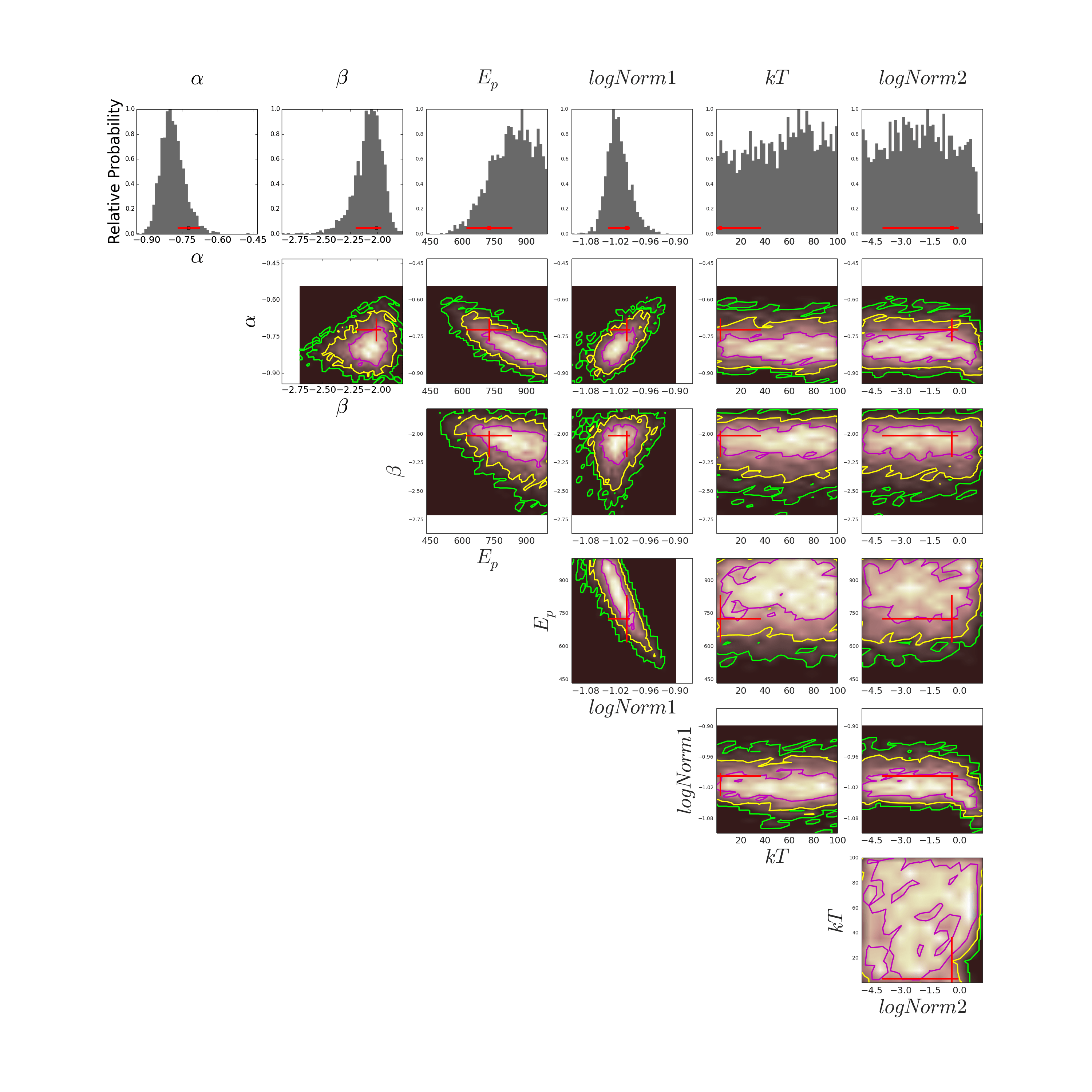}};
\node (fig2) at (11,8)
 {\includegraphics[width=2.0in]{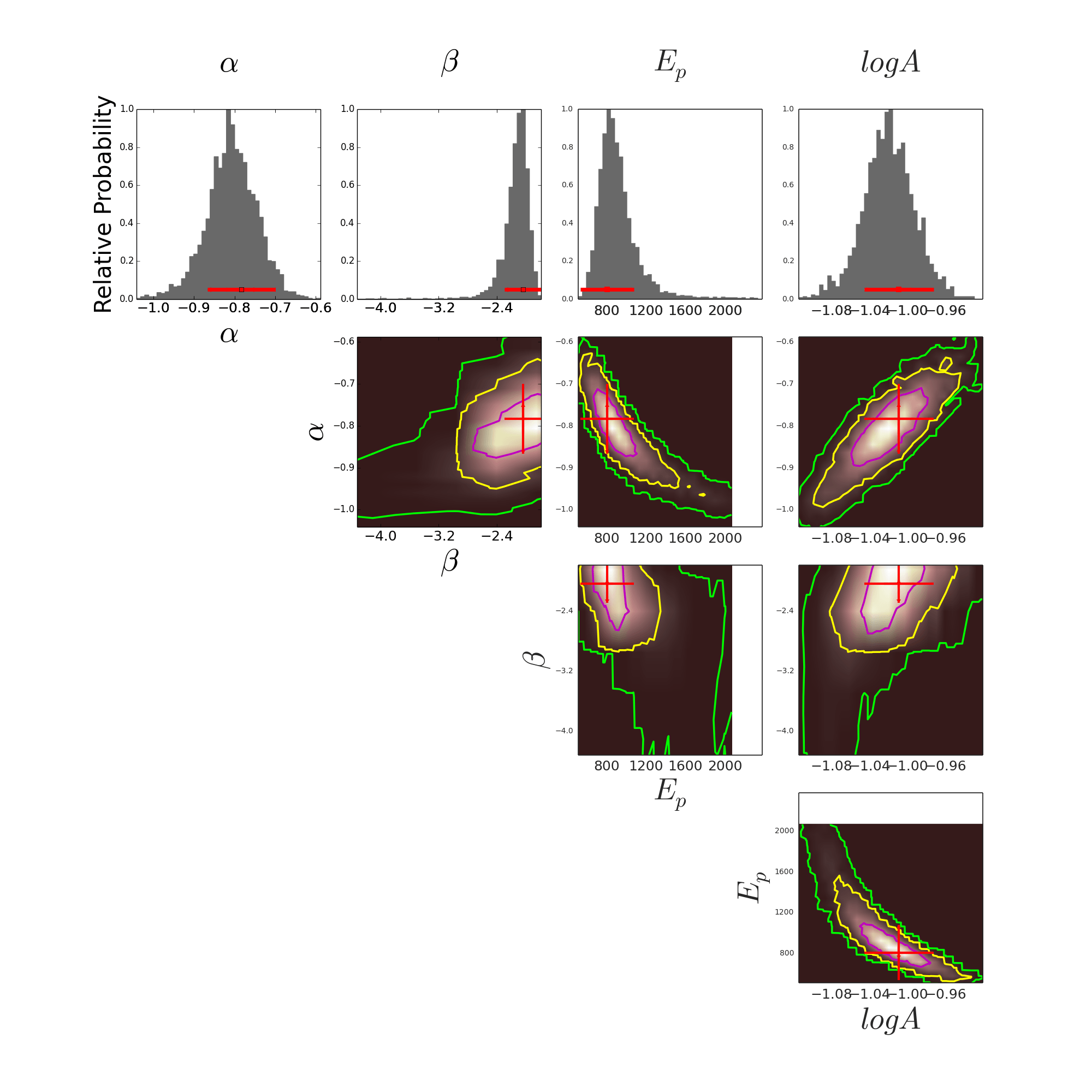}};
\end{tikzpicture}
 \caption{An example showing the effect of adding a blackbody component to the Band function model when fitting to the data in the time bin between 187.23 s and 187.67 s. {\it Bottom Left}: the likelihood map of the parameter-constraint outputs for the Band fit, where all the parameters are constrained. The goodness of fit, PGSTAT/dof, as shown in Supplementary Table~\ref{tab:specfitting}, is 335.55/360. {\it Upper Right}: the likelihood map of the parameter-constraint outputs for the Band+blackbody fit. The PGSTAT/dof is 334.01/358. According to the Bayesian information criterion (BIC\cite{S21}), such value of goodness does not imply any improvement in comparison with the Band-only fit. More importantly, we noticed the flat-shape of likelihood map of the two blackbody parameters (i.e, blackbody temperate kT and normalization parameter logNorm2), implying they are unconstrained. The fluence of the such blackbody component is only $\sim$ 0.05\% of the total fluence. The above facts suggest that the blackbody component is not required to fit the observational data. }
\label{fig:addingbb}
\end{figure*}

\begin{figure*} \centering
\begin{tikzpicture}[
 every node/.style={anchor=south west,inner sep=0pt},
 x=1mm, y=1mm,
 ] 
 \node (fig1) at (0,0)
 {\includegraphics[width=7.0in]{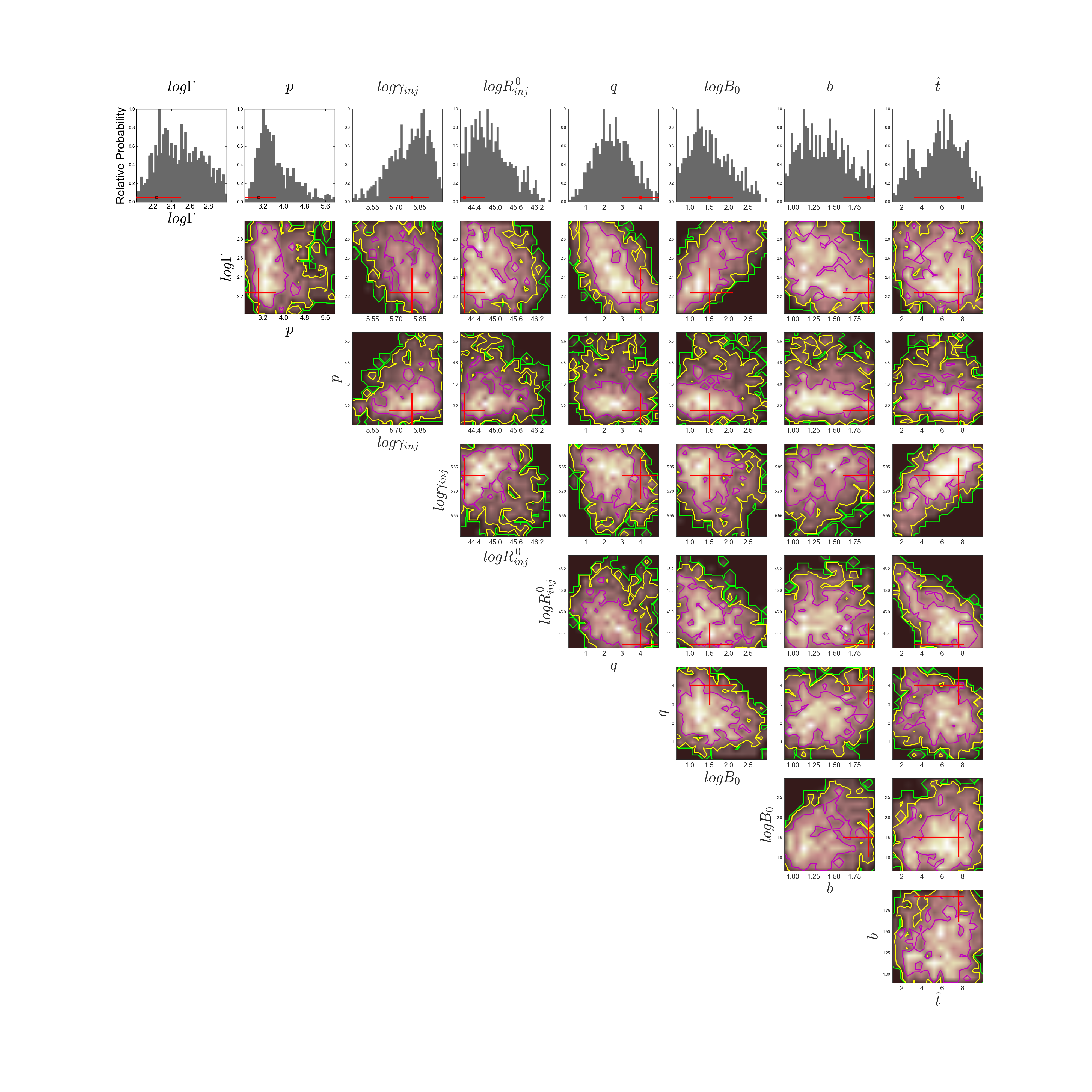}};
\node (fig2) at (12,8)
 {\includegraphics[width=2.3in]{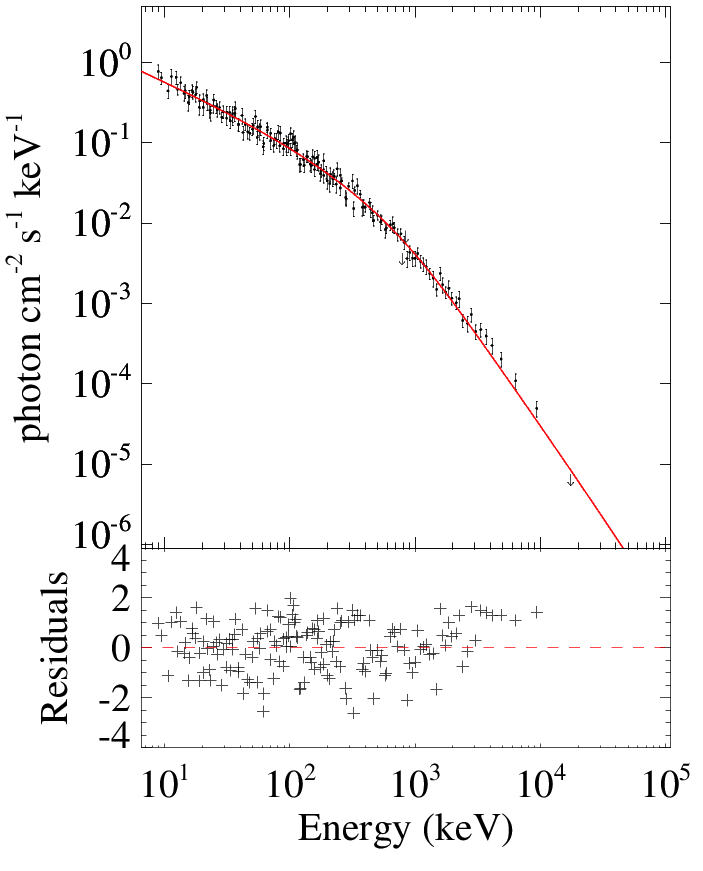}};
\end{tikzpicture}
 \caption{An example fit to the data in one time bin (between 187.23 s and 187.67 s) using the synchrotron model (bottom-left) and constraints on the model parameters (upper-right). Histograms and contours show the likelihood map of the parameter-constraint outputs from our McSpecFit package. Red crosses mark the best-fit values and 1$\sigma$ error bars. All the constrained parameters fall into the reasonable ranges for the synchrotron theoretical model. A Band function can equally fit to the data with $\alpha$=-0.8 and $\beta$=-2.0 Supplementary Table \ref{tab:specfitting}).}
\label{fig:syncfit}
\end{figure*}

\begin{figure*}

\begin{tabular}{ccc}
\includegraphics[keepaspectratio,clip,width=0.32\textwidth]{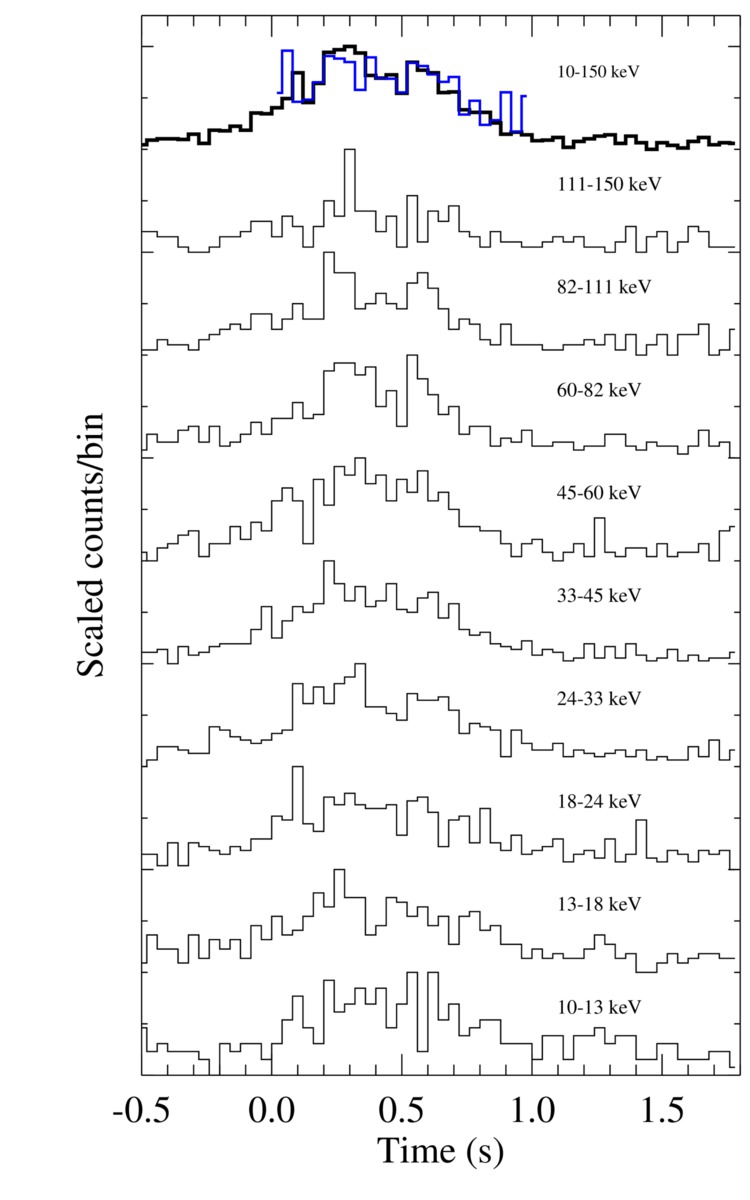} &
\includegraphics[keepaspectratio,clip,width=0.32\textwidth]{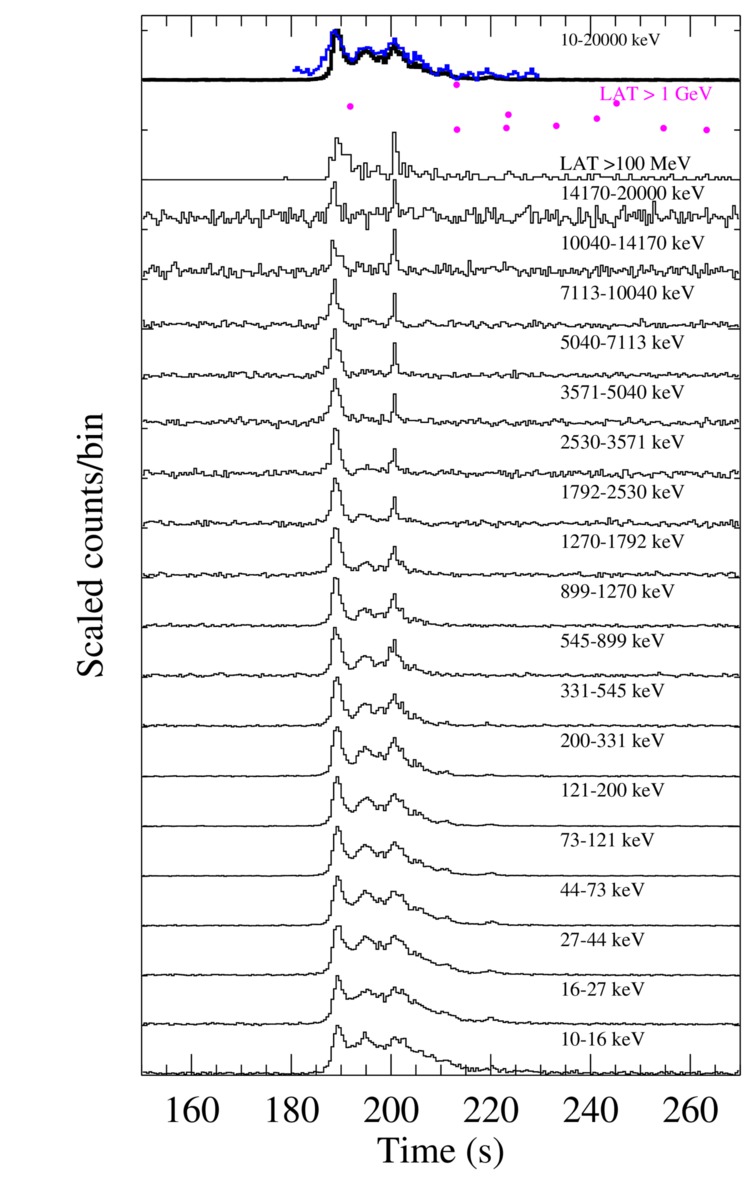} &
\includegraphics[keepaspectratio,clip,width=0.32\textwidth]{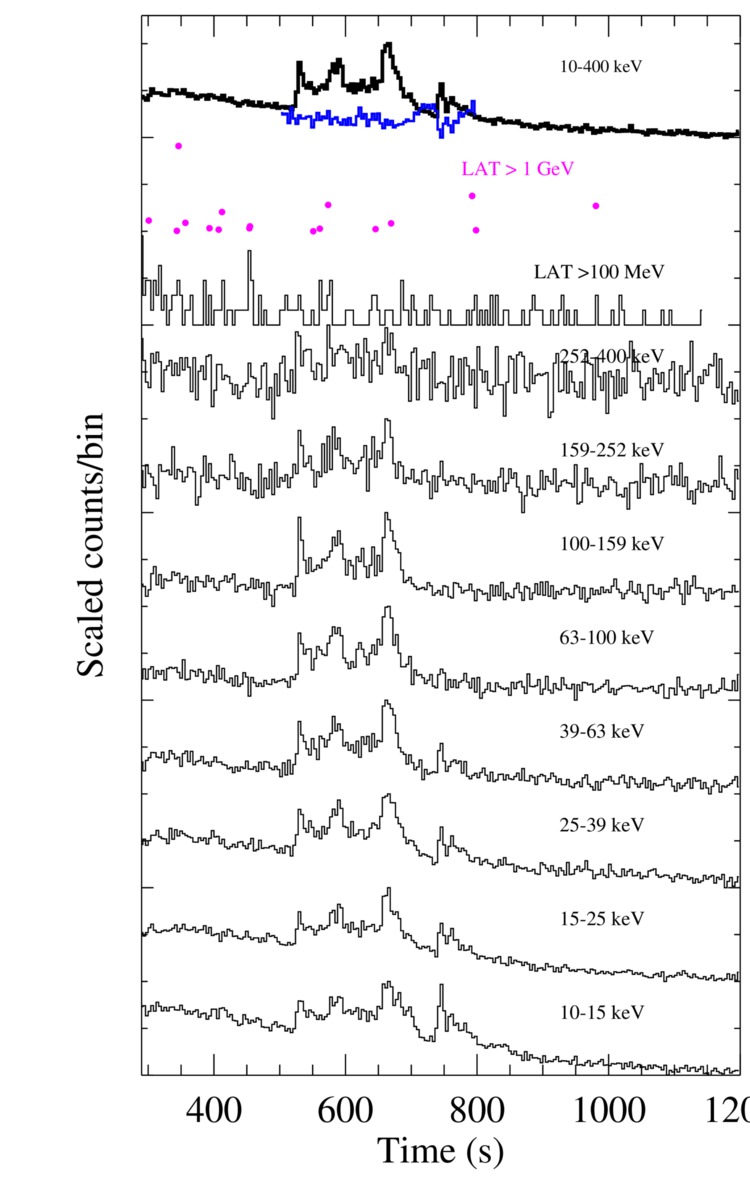} \\

\end{tabular}
\caption{ 
Energy-dependent light curves of the three sub-bursts. In each plot, the full-energy-range light curves are plotted at the top with the thick black curves, while the ligh tcurves in each energy bin are plotted beneath. Also the hardness ratios (defined as HR$=\frac{{\rm Rate}_{E_2,E_3}}{{\rm Rate}_{E_1,E_2}}$, where $E_1$ and $E_3$ define the lower and higher end of the energy range of the selected NaI detectors, and $E_2=\sqrt{E_1\times E_3}$) are plotted in solid blue curves. The LAT-band photons are plotted individually for the 2nd and 3rd sub-bursts. 
}
\label{fig:wave}
\end{figure*}

\begin{figure*}

\begin{tabular}{ccc}
\includegraphics[keepaspectratio,clip,width=0.33\textwidth]{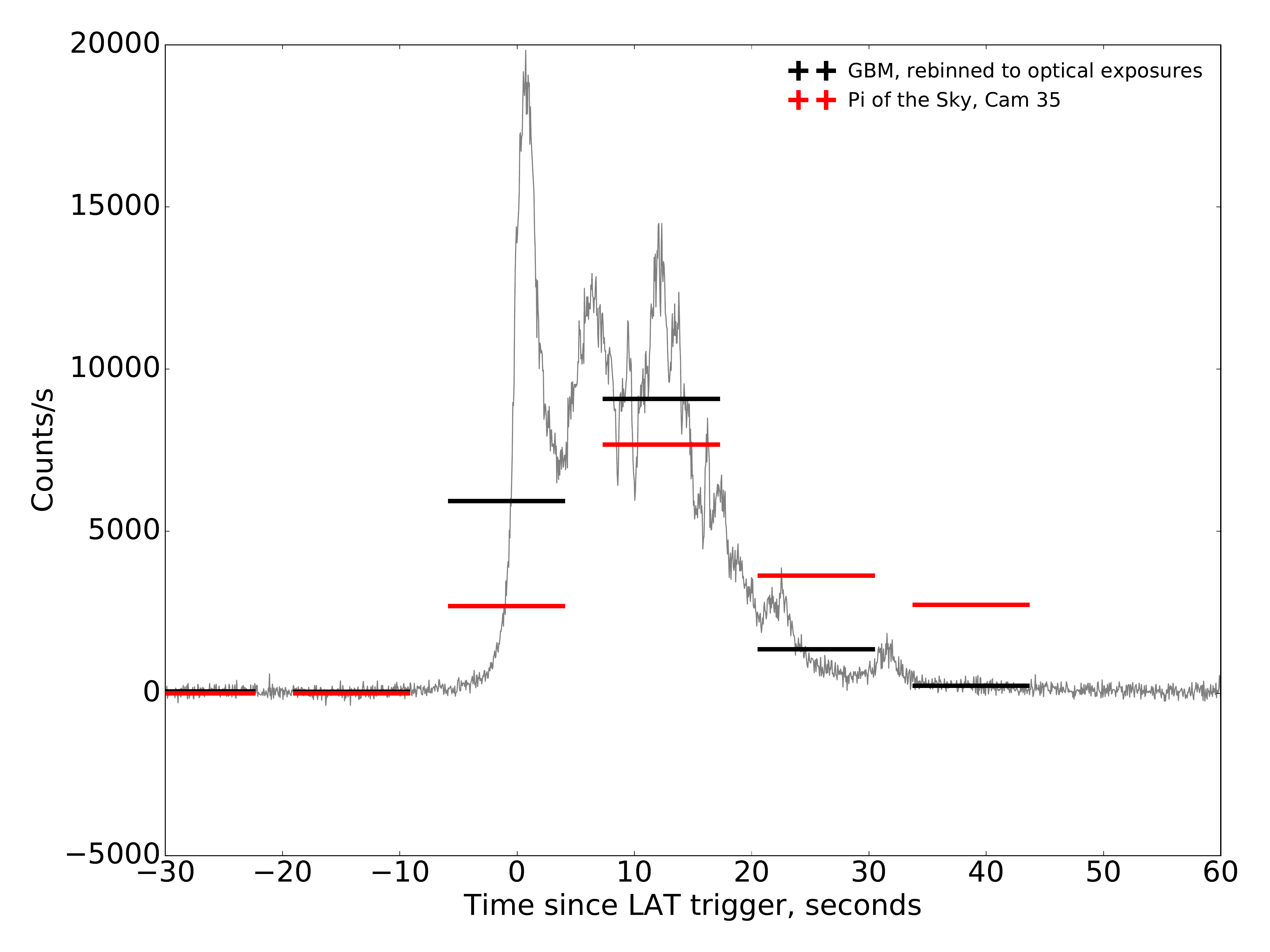}&
\includegraphics[keepaspectratio,clip,width=0.33\textwidth]{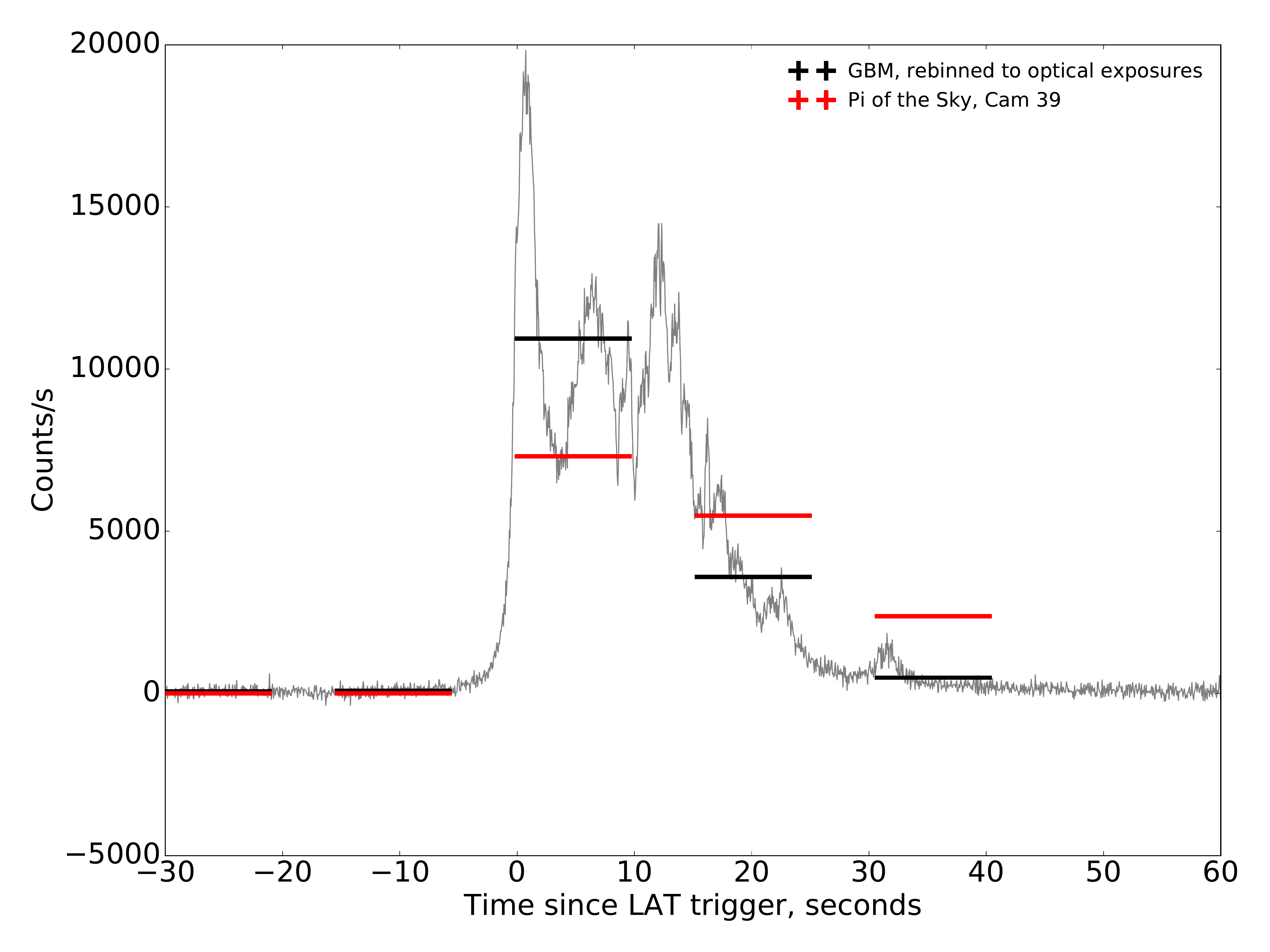}&
\includegraphics[keepaspectratio,clip,width=0.33\textwidth]{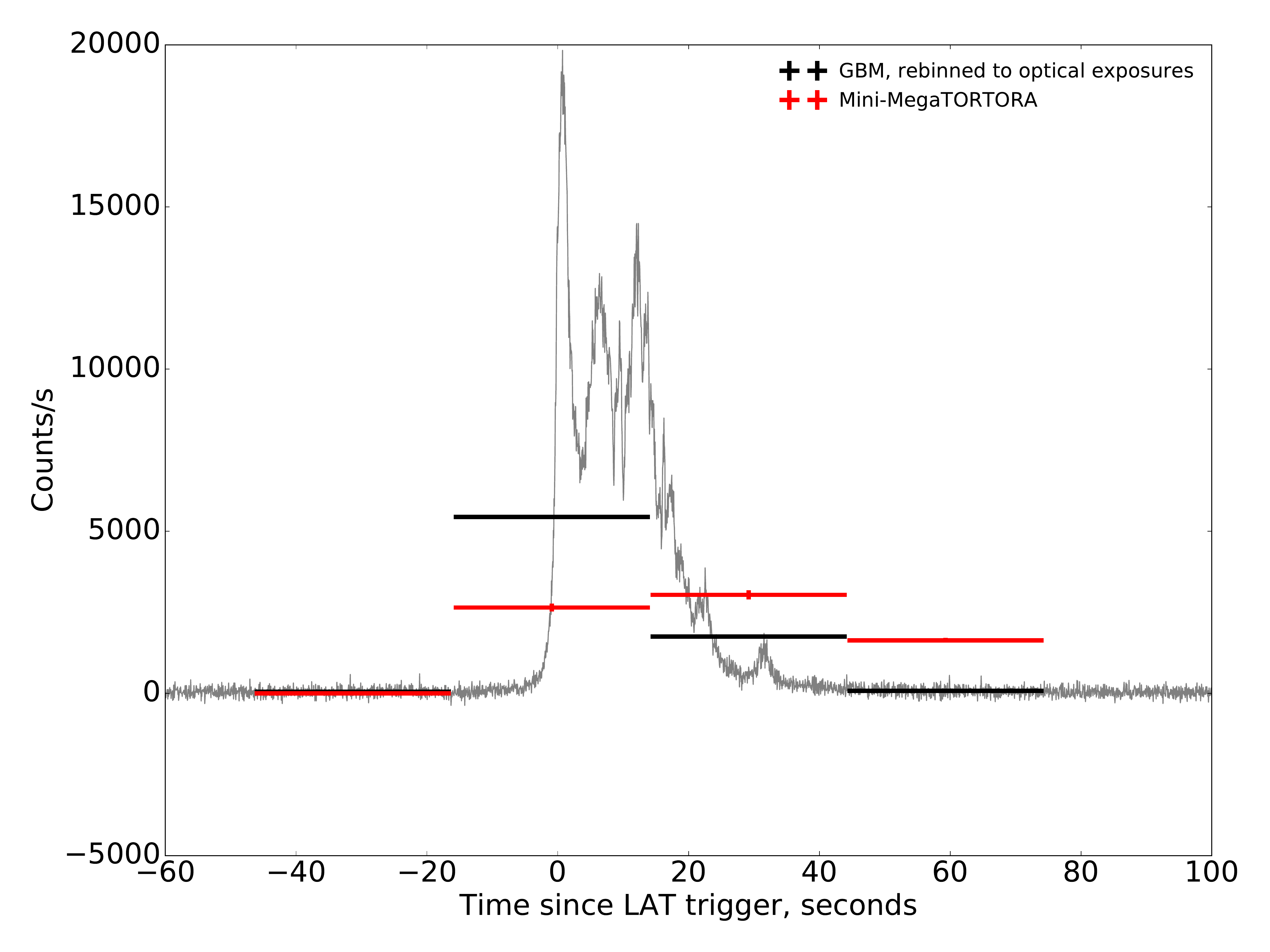}\\

\end{tabular}
\caption{ 
Correlations between GBM prompt $\gamma$-ray (black; 15-350 keV ) and optical observations (red) made by Pi of the Sky (Cam 35), Pi of the Sky (Cam 39) and Mini-Mega TORTORA respectively. Lags are calcualted between the optical flux and the averaged GBM count rate in the optical exposure intervals. The values of lag are listed in Table \ref{tab:lag_opt_gbm}.}
\label{fig:opt_mev_lag}
\end{figure*}

\begin{figure*}

\begin{tabular}{c}
\includegraphics[keepaspectratio,clip,width=0.99\textwidth]{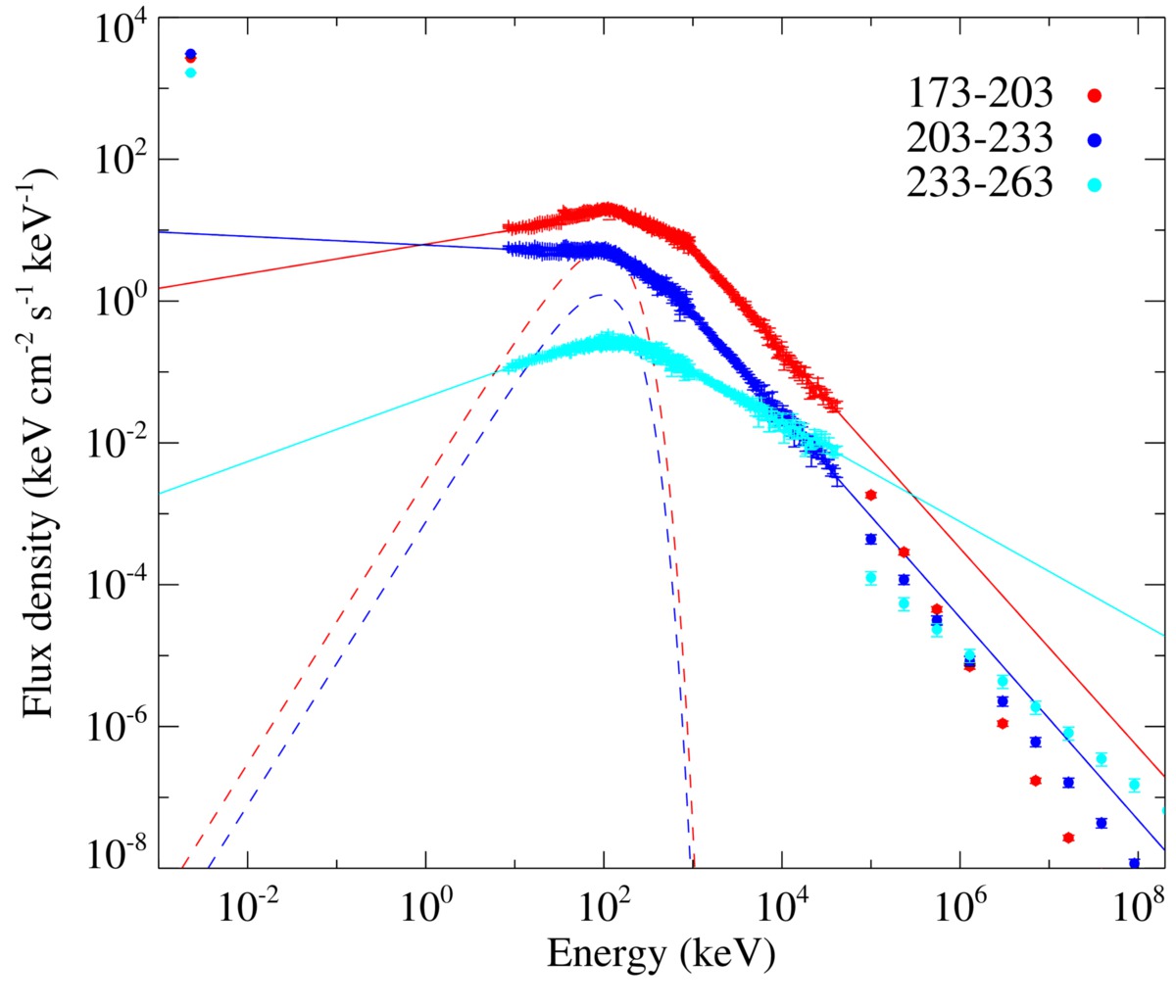} \\

\end{tabular}
\caption{ 
The spectral energy distribution (SED) from optical to LAT energies in each optical observational bins during the second sub-burst. The model curves (solid lines) are derived using GBM data only, which are extrapolated to the optical and LAT energy bands. Such model curves are only rough estimates of the true SEDs due to the large time bins used and the goodness of those fits is not statistically favorable (Table \ref{tab:sed}). Adding an additional blackbody component (dashed) significantly improves the goodness of the fits from the Band-only case (Table \ref{tab:sed}). However, such an apparent blackbody component is an effect of spectral evolution, since in finer spectral bins, no blackbody component is needed (Figure 1 and Supplementary Table \ref{tab:specfitting}). 
}
\label{fig:sed}
\end{figure*}

\clearpage


\begin{table}

\caption{Time-resolved Spectral Fitting (Sub-burst I)}
\setlength{\tabcolsep}{2pt}

\begin{tiny}
\begin{center}
\setlength\tabcolsep{1.0pt} 
\begin{tabular}{l|ll|lll|ll|llll}
\hline

 \multicolumn{1}{c|}{Tim} & \multicolumn{2}{c|}{Power-Law Function Fitting} & \multicolumn{3}{c|}{Cutoff Power-Law Fitting} & \multicolumn{2}{c|}{ Blackbody Fitting} & \multicolumn{4}{c}{ Multi-Color Blackbody Fitting} \\
\hline
 t1 $\sim$ t2 (s) & $\alpha$ & PGSTAT/dof & $\alpha$ & E$_p$ (keV) & PGSTAT/dof & kT (keV) & PGSTAT/dof & $kT_{\min}$ (keV) & $kT_{\max}$ (keV) & $m$ & PGSTAT/dof \\ 
\hline
-0.30 $\sim$ -0.08 &$-2.11_{-0.26}^{+0.16}$&210.07/362& & unconstrained & &$9.06_{-1.24}^{+2.33}$&203.42/362&$6.29_{-1.72}^{+49.66}$&$36.28_{-26.30}^{+707.87}$&$-1.85_{-0.04}^{+2.15}$&202.15/360\\
-0.08 $\sim$ 0.13 &$-1.93_{-0.05}^{+0.05}$&301.94/362&$0.02_{-0.24}^{+0.37}$&$57.64_{-3.59}^{+4.14}$&207.89/361&$13.64_{-0.67}^{+0.63}$&222.68/362&$4.10_{-0.55}^{+1.85}$&$25.10_{-2.16}^{+11.82}$&$0.48_{-0.95}^{+0.95}$&207.83/360\\
0.13 $\sim$ 0.35 &$-1.86_{-0.03}^{+0.03}$&460.23/362&$-0.16_{-0.18}^{+0.08}$&$70.41_{-1.68}^{+4.43}$&215.74/361&$15.59_{-0.46}^{+0.50}$&278.12/362&$3.69_{-0.03}^{+4.04}$&$33.10_{-1.06}^{+11.46}$&$0.25_{-1.26}^{+0.04}$&215.28/360\\
0.35 $\sim$ 0.57 &$-1.87_{-0.04}^{+0.03}$&395.11/362&$0.04_{-0.12}^{+0.08}$&$65.45_{-1.32}^{+5.19}$&199.27/361&$15.38_{-0.45}^{+0.51}$&228.15/362&$2.24_{-0.36}^{+2.33}$&$26.83_{-1.03}^{+9.45}$&$0.72_{-0.60}^{+0.40}$&198.74/360\\
0.57 $\sim$ 0.78 &$-1.89_{-0.04}^{+0.04}$&341.40/362&$0.24_{-0.24}^{+0.27}$&$61.23_{-2.85}^{+3.23}$&192.97/361&$14.60_{-0.54}^{+0.57}$&209.76/362&$3.37_{-0.58}^{+2.48}$&$23.54_{-0.35}^{+10.25}$&$0.82_{-0.88}^{+0.79}$&192.77/360\\
0.78 $\sim$ 1.00 &$-2.30_{-0.14}^{+0.11}$&233.36/362& & unconstrained & &$8.47_{-0.77}^{+1.08}$&217.85/362&$3.21_{-1.49}^{+26.59}$&$19.12_{-10.13}^{+279.71}$&$-0.38_{-1.48}^{+1.13}$&212.94/360\\
\hline
\end{tabular}
 \end{center}
\end{tiny}
\label{tab:specfitting}
\end{table}

\setcounter{table}{0}
\begin{table}
\caption{Time-resolved Spectral Fitting (Sub-burst III)}
\setlength{\tabcolsep}{2pt}

\begin{tiny}
\begin{center}
\begin{tabular}{l|ll|lll}
\hline

 \multicolumn{1}{c|}{Time} & \multicolumn{2}{c|}{Power-Law Function Fitting} & \multicolumn{3}{c}{Cutoff Power-Law Fitting} \\
\hline
 t1 $\sim$ t2 (s) & $\alpha$ & PGSTAT/dof & $\alpha$ & E$_p$ (keV) & PGSTAT/dof \\ 
\hline

520.00 $\sim$ 535.00 &$-1.46_{-0.03}^{+0.02}$&349.13/362& & unconstrained &\\
535.00 $\sim$ 542.00 &$-1.57_{-0.06}^{+0.04}$&320.62/362& & unconstrained &\\
542.00 $\sim$ 559.00 &$-1.54_{-0.05}^{+0.03}$&403.90/362& & unconstrained &\\
570.00 $\sim$ 578.00 &$-1.54_{-0.04}^{+0.03}$&366.79/362& & unconstrained &\\
578.00 $\sim$ 585.00 &$-1.57_{-0.03}^{+0.03}$&425.14/362& & unconstrained &\\
585.00 $\sim$ 593.00 &$-1.61_{-0.03}^{+0.02}$&415.58/362&$-1.20_{-0.13}^{+0.07}$&$188.61_{-19.34}^{+70.23}$&384.23/361\\
593.00 $\sim$ 600.00 &$-1.63_{-0.05}^{+0.05}$&413.41/362& & unconstrained &\\
650.00 $\sim$ 658.00 &$-1.61_{-0.04}^{+0.03}$&353.24/362& & unconstrained &\\
658.00 $\sim$ 661.50 &$-1.71_{-0.04}^{+0.03}$&426.62/362&$-1.40_{-0.21}^{+0.06}$&$171.80_{-16.82}^{+262.90}$&411.65/361\\
661.50 $\sim$ 665.00 &$-1.64_{-0.03}^{+0.02}$&411.45/362& & unconstrained &\\
665.00 $\sim$ 669.00 &$-1.66_{-0.03}^{+0.03}$&437.81/362& & unconstrained &\\
669.00 $\sim$ 673.00 &$-1.65_{-0.04}^{+0.03}$&386.22/362& & unconstrained &\\
673.00 $\sim$ 680.00 &$-1.68_{-0.04}^{+0.03}$&399.92/362& & unconstrained &\\
740.00 $\sim$ 750.00 &$-2.29_{-0.08}^{+0.07}$&397.40/362& & unconstrained &\\
750.00 $\sim$ 760.00 &$-2.73_{-0.29}^{+0.19}$&318.07/362& & unconstrained &\\
760.00 $\sim$ 770.00 &$-2.33_{-0.13}^{+0.12}$&379.96/362& & unconstrained &\\
770.00 $\sim$ 780.00 &$-2.38_{-0.33}^{+0.25}$&319.50/362& & unconstrained &\\

\hline
\end{tabular}
 \end{center}
\end{tiny}
\end{table}

\setcounter{table}{0}
\begin{sidewaystable}

\caption{Time-resolved Spectral Fitting (Sub-burst II) }
\setlength{\tabcolsep}{1.5pt}

\begin{tiny}
\begin{center}

\begin{tabular}{l|llll|lllllllll}
\hline
 \multicolumn{1}{c|}{Time} & \multicolumn{4}{c|}{Band Function Fitting} & \multicolumn{9}{c}{Synchrotron Model Fitting} \\
\hline
 t1 $\sim$ t2 (s) & $\alpha$ & $\beta$ & $E_p$ (keV) & PGSTAT/dof & log $\Gamma$ & $p$ & log $\gamma_{inj}$ & log R$_{inj}^0$ (s$^{-1}$) & q & log B$_0$ (G) & b & $\hat{t}$ (s) & PGSTAT/dof \\ 
\hline
\hline

187.23 $\sim$ 187.67 &$-0.78_{-0.09}^{+0.03}$& $-2.04_{-0.25}^{+0.06}$&$796.76_{-53.56}^{+272.28}$& 335.56/360&$2.24_{-0.22}^{+0.26}$&$3.04_{-0.54}^{+0.67}$&$5.80_{-0.15}^{+0.11}$&$44.10_{-0.10}^{+0.59}$&$4.02_{-1.05}^{+0.98}$&$1.52_{-0.50}^{+0.60}$&$1.92_{-0.31}^{+0.08}$&$7.62_{-4.39}^{+0.49}$&332.84/356\\
187.67 $\sim$ 188.04 &$-0.70_{-0.06}^{+0.04}$& $-1.97_{-0.11}^{+0.05}$&$747.31_{-55.80}^{+151.91}$& 385.59/360&$2.74_{-0.43}^{+0.03}$&$3.06_{-0.08}^{+0.36}$&$5.56_{-0.06}^{+0.12}$&$45.68_{-0.62}^{+0.62}$&$4.09_{-1.75}^{+0.02}$&$1.18_{-0.31}^{+0.53}$&$0.97_{-0.07}^{+0.28}$&$2.64_{-0.98}^{+1.83}$&377.36/356\\
188.04 $\sim$ 188.16 &$-0.74_{-0.10}^{+0.03}$& $-2.12_{-0.49}^{+0.09}$&$909.67_{-72.76}^{+429.38}$& 337.59/360&$2.70_{-0.47}^{+0.06}$&$3.48_{-0.15}^{+1.04}$&$5.58_{-0.09}^{+0.11}$&$44.95_{-0.52}^{+0.74}$&$4.52_{-1.07}^{+0.45}$&$1.87_{-0.96}^{+0.16}$&$1.26_{-0.19}^{+0.39}$&$3.64_{-2.24}^{+2.27}$&329.63/356\\
188.16 $\sim$ 188.33 &$-0.71_{-0.05}^{+0.04}$& $-2.19_{-0.15}^{+0.10}$&$997.93_{-98.30}^{+169.32}$& 383.22/360&$2.65_{-0.29}^{+0.20}$&$3.51_{-0.40}^{+0.40}$&$5.77_{-0.08}^{+0.08}$&$44.12_{-0.11}^{+0.59}$&$4.21_{-0.82}^{+0.78}$&$1.16_{-0.19}^{+0.92}$&$0.91_{-0.01}^{+0.25}$&$7.07_{-2.33}^{+1.00}$&369.83/356\\
188.33 $\sim$ 188.43 &$-0.63_{-0.04}^{+0.06}$& $-2.23_{-0.17}^{+0.10}$&$809.54_{-98.03}^{+96.38}$& 353.35/360&$2.60_{-0.15}^{+0.24}$&$4.40_{-0.29}^{+0.60}$&$5.91_{-0.09}^{+0.06}$&$45.77_{-0.65}^{+0.65}$&$4.73_{-0.73}^{+0.24}$&$1.12_{-0.10}^{+0.90}$&$1.33_{-0.26}^{+0.26}$&$4.01_{-1.67}^{+1.67}$&351.01/356\\
188.43 $\sim$ 188.56 &$-0.61_{-0.05}^{+0.04}$& $-2.14_{-0.11}^{+0.06}$&$816.62_{-62.76}^{+97.11}$& 416.56/360&$2.44_{-0.19}^{+0.19}$&$3.96_{-0.16}^{+0.45}$&$5.99_{-0.08}^{+0.01}$&$44.86_{-0.53}^{+0.70}$&$4.34_{-1.11}^{+0.17}$&$0.86_{-0.18}^{+0.56}$&$1.19_{-0.15}^{+0.44}$&$7.25_{-2.79}^{+0.72}$&415.50/356\\
188.56 $\sim$ 188.76 &$-0.62_{-0.03}^{+0.03}$& $-2.14_{-0.07}^{+0.05}$&$783.85_{-45.84}^{+70.51}$& 385.27/360&$2.80_{-0.24}^{+0.10}$&$3.80_{-0.11}^{+0.23}$&$5.89_{-0.04}^{+0.08}$&$44.06_{-0.06}^{+0.51}$&$4.71_{-0.55}^{+0.28}$&$1.07_{-0.10}^{+1.07}$&$0.94_{-0.04}^{+0.27}$&$7.64_{-2.35}^{+0.34}$&384.92/356\\
188.76 $\sim$ 188.96 &$-0.63_{-0.04}^{+0.03}$& $-2.13_{-0.07}^{+0.05}$&$706.47_{-44.51}^{+62.60}$& 396.40/360&$2.79_{-0.28}^{+0.09}$&$3.82_{-0.14}^{+0.21}$&$5.98_{-0.07}^{+0.02}$&$44.36_{-0.04}^{+1.03}$&$4.66_{-0.63}^{+0.34}$&$1.66_{-0.74}^{+0.32}$&$1.38_{-0.36}^{+0.15}$&$7.72_{-2.39}^{+0.69}$&389.92/356\\
188.96 $\sim$ 189.00 &$-0.57_{-0.11}^{+0.08}$& $-2.13_{-0.32}^{+0.08}$&$580.64_{-78.33}^{+185.22}$& 337.52/360&$2.81_{-0.42}^{+0.04}$&$4.06_{-0.18}^{+0.79}$&$5.96_{-0.10}^{+0.04}$&$44.28_{-0.28}^{+0.66}$&$4.69_{-0.75}^{+0.31}$&$1.24_{-0.22}^{+0.97}$&$1.14_{-0.07}^{+0.55}$&$7.89_{-3.19}^{+0.62}$&336.31/356\\
189.00 $\sim$ 189.06 &$-0.64_{-0.08}^{+0.04}$& $-2.34_{-0.31}^{+0.10}$&$725.28_{-57.78}^{+171.55}$& 413.43/360&$2.61_{-0.22}^{+0.24}$&$4.78_{-0.39}^{+0.67}$&$5.80_{-0.02}^{+0.14}$&$45.86_{-0.64}^{+0.64}$&$3.62_{-0.81}^{+0.89}$&$0.89_{-0.22}^{+0.57}$&$0.95_{-0.05}^{+0.24}$&$4.13_{-1.56}^{+1.56}$&405.71/356\\
189.06 $\sim$ 189.11 &$-0.56_{-0.07}^{+0.05}$& $-2.30_{-0.20}^{+0.10}$&$598.22_{-44.98}^{+88.30}$& 364.04/360&$2.90_{-0.20}^{+0.10}$&$4.58_{-0.34}^{+0.73}$&$5.96_{-0.09}^{+0.04}$&$44.51_{-0.10}^{+1.26}$&$4.61_{-0.73}^{+0.38}$&$2.42_{-0.65}^{+0.54}$&$1.66_{-0.57}^{+0.06}$&$6.99_{-2.40}^{+1.22}$&370.15/356\\
189.11 $\sim$ 189.16 &$-0.64_{-0.08}^{+0.04}$& $-2.38_{-0.31}^{+0.10}$&$635.67_{-46.62}^{+122.93}$& 375.07/360&$2.60_{-0.17}^{+0.26}$&$4.51_{-0.31}^{+0.79}$&$5.93_{-0.19}^{+0.03}$&$45.04_{-0.65}^{+0.58}$&$4.49_{-1.34}^{+0.09}$&$0.87_{-0.20}^{+0.48}$&$1.09_{-0.06}^{+0.56}$&$6.36_{-1.46}^{+2.24}$&373.65/356\\
189.16 $\sim$ 189.37 &$-0.60_{-0.03}^{+0.03}$& $-2.22_{-0.08}^{+0.05}$&$600.37_{-34.89}^{+46.93}$& 395.95/360&$2.65_{-0.15}^{+0.19}$&$4.27_{-0.11}^{+0.36}$&$5.99_{-0.08}^{+0.01}$&$45.05_{-0.58}^{+0.65}$&$4.80_{-0.55}^{+0.19}$&$1.46_{-0.55}^{+0.30}$&$1.53_{-0.21}^{+0.21}$&$6.28_{-1.70}^{+1.56}$&398.70/356\\
189.37 $\sim$ 189.48 &$-0.60_{-0.07}^{+0.02}$& $-2.15_{-0.11}^{+0.06}$&$580.87_{-30.75}^{+91.50}$& 393.13/360&$2.39_{-0.03}^{+0.38}$&$3.93_{-0.07}^{+0.53}$&$5.88_{-0.09}^{+0.07}$&$44.47_{-0.04}^{+1.18}$&$4.45_{-1.46}^{+0.06}$&$0.76_{-0.09}^{+0.60}$&$1.00_{-0.10}^{+0.33}$&$8.51_{-3.12}^{+0.03}$&385.10/356\\
189.47 $\sim$ 189.58 &$-0.57_{-0.07}^{+0.03}$& $-2.16_{-0.13}^{+0.04}$&$508.10_{-25.51}^{+72.54}$& 410.90/360&$2.73_{-0.28}^{+0.15}$&$4.18_{-0.13}^{+0.51}$&$5.72_{-0.09}^{+0.09}$&$44.07_{-0.07}^{+0.64}$&$4.80_{-0.74}^{+0.18}$&$1.22_{-0.28}^{+0.87}$&$0.96_{-0.05}^{+0.29}$&$7.20_{-1.81}^{+1.41}$&410.50/356\\
189.58 $\sim$ 189.78 &$-0.54_{-0.05}^{+0.02}$& $-2.15_{-0.08}^{+0.04}$&$481.26_{-17.76}^{+43.45}$& 400.01/360&$2.79_{-0.28}^{+0.07}$&$4.11_{-0.16}^{+0.23}$&$5.98_{-0.07}^{+0.02}$&$44.61_{-0.38}^{+0.74}$&$4.95_{-0.48}^{+0.05}$&$1.08_{-0.19}^{+0.96}$&$1.17_{-0.14}^{+0.48}$&$6.96_{-1.33}^{+1.69}$&426.18/356\\
189.78 $\sim$ 189.98 &$-0.62_{-0.04}^{+0.04}$& $-2.16_{-0.08}^{+0.05}$&$517.74_{-34.01}^{+43.86}$& 396.70/360&$2.55_{-0.11}^{+0.28}$&$3.93_{-0.09}^{+0.33}$&$5.98_{-0.09}^{+0.02}$&$44.91_{-0.50}^{+0.52}$&$4.87_{-0.55}^{+0.13}$&$0.74_{-0.07}^{+0.53}$&$1.18_{-0.10}^{+0.47}$&$6.80_{-1.34}^{+1.59}$&395.21/356\\
189.98 $\sim$ 190.04 &$-0.62_{-0.08}^{+0.10}$& $-2.14_{-0.18}^{+0.10}$&$404.09_{-54.17}^{+72.15}$& 290.77/360&$2.60_{-0.25}^{+0.21}$&$3.78_{-0.14}^{+0.78}$&$5.79_{-0.12}^{+0.12}$&$44.15_{-0.15}^{+0.70}$&$4.73_{-1.01}^{+0.26}$&$1.69_{-0.74}^{+0.39}$&$1.51_{-0.35}^{+0.21}$&$8.21_{-2.70}^{+0.61}$&290.87/356\\
190.04 $\sim$ 190.33 &$-0.66_{-0.04}^{+0.04}$& $-2.14_{-0.08}^{+0.05}$&$474.23_{-30.55}^{+40.49}$& 407.95/360&$2.57_{-0.14}^{+0.26}$&$3.79_{-0.14}^{+0.19}$&$5.86_{-0.18}^{+0.05}$&$44.81_{-0.38}^{+0.86}$&$4.92_{-0.76}^{+0.06}$&$1.23_{-0.35}^{+0.62}$&$1.41_{-0.33}^{+0.19}$&$6.30_{-1.53}^{+1.59}$&401.00/356\\
190.33 $\sim$ 190.66 &$-0.65_{-0.04}^{+0.04}$& $-2.17_{-0.10}^{+0.05}$&$411.31_{-24.96}^{+35.18}$& 424.03/360&$2.78_{-0.41}^{+0.06}$&$4.00_{-0.16}^{+0.30}$&$5.94_{-0.12}^{+0.05}$&$44.40_{-0.06}^{+0.94}$&$4.91_{-0.69}^{+0.09}$&$1.51_{-0.61}^{+0.21}$&$1.44_{-0.37}^{+0.22}$&$7.26_{-1.77}^{+1.11}$&417.86/356\\
190.66 $\sim$ 190.96 &$-0.70_{-0.05}^{+0.05}$& $-2.40_{-0.21}^{+0.12}$&$426.75_{-36.08}^{+41.63}$& 377.34/360&$2.62_{-0.21}^{+0.22}$&$4.53_{-0.13}^{+0.74}$&$5.88_{-0.15}^{+0.06}$&$44.57_{-0.18}^{+0.93}$&$4.79_{-0.81}^{+0.19}$&$1.04_{-0.15}^{+0.78}$&$1.25_{-0.14}^{+0.38}$&$6.78_{-1.11}^{+1.79}$&369.82/356\\
190.96 $\sim$ 191.47 &$-0.63_{-0.05}^{+0.03}$& $-2.20_{-0.10}^{+0.05}$&$326.57_{-16.83}^{+30.33}$& 365.23/360&$2.66_{-0.21}^{+0.20}$&$4.28_{-0.25}^{+0.29}$&$5.91_{-0.13}^{+0.09}$&$44.82_{-0.53}^{+0.32}$&$4.46_{-1.21}^{+0.07}$&$0.92_{-0.24}^{+0.59}$&$1.24_{-0.09}^{+0.54}$&$6.71_{-1.02}^{+1.66}$&362.36/356\\
191.47 $\sim$ 192.00 &$-0.67_{-0.06}^{+0.02}$& $-2.37_{-0.15}^{+0.06}$&$315.48_{-10.64}^{+32.05}$& 362.05/360&$2.78_{-0.41}^{+0.06}$&$4.69_{-0.19}^{+0.59}$&$5.65_{-0.00}^{+0.26}$&$44.62_{-0.21}^{+0.89}$&$4.24_{-0.64}^{+0.64}$&$0.95_{-0.03}^{+1.27}$&$0.92_{-0.02}^{+0.26}$&$5.66_{-0.29}^{+3.00}$&351.32/356\\
192.00 $\sim$ 192.47 &$-0.68_{-0.04}^{+0.05}$& $-2.32_{-0.14}^{+0.08}$&$287.35_{-17.68}^{+22.16}$& 378.14/360&$2.91_{-0.21}^{+0.09}$&$4.51_{-0.20}^{+0.68}$&$5.88_{-0.23}^{+0.03}$&$44.46_{-0.18}^{+0.96}$&$4.53_{-0.78}^{+0.46}$&$1.66_{-0.71}^{+0.35}$&$1.43_{-0.18}^{+0.35}$&$6.85_{-1.03}^{+1.87}$&377.34/356\\
192.47 $\sim$ 192.85 &$-0.69_{-0.04}^{+0.04}$& $-2.51_{-6.45}^{+0.29}$&$327.30_{-29.49}^{+29.49}$& 326.54/360&$2.58_{-0.16}^{+0.27}$&$5.16_{-0.31}^{+0.54}$&$5.88_{-0.17}^{+0.05}$&$44.57_{-0.27}^{+0.89}$&$4.14_{-1.51}^{+0.05}$&$1.24_{-0.31}^{+0.83}$&$1.43_{-0.26}^{+0.34}$&$7.78_{-1.83}^{+1.10}$&318.67/356\\
192.85 $\sim$ 193.19 &$-0.60_{-0.05}^{+0.05}$& $-2.33_{-0.14}^{+0.07}$&$296.03_{-16.60}^{+26.59}$& 384.53/360&$2.66_{-0.29}^{+0.04}$&$4.99_{-0.33}^{+0.47}$&$5.78_{-0.08}^{+0.14}$&$44.96_{-0.61}^{+0.46}$&$4.76_{-0.66}^{+0.24}$&$1.29_{-0.42}^{+0.43}$&$1.40_{-0.16}^{+0.34}$&$5.46_{-0.17}^{+3.22}$&388.92/356\\
193.19 $\sim$ 193.74 &$-0.61_{-0.03}^{+0.03}$& $-2.49_{-0.17}^{+0.05}$&$371.57_{-15.06}^{+20.17}$& 412.83/360&$2.72_{-0.20}^{+0.13}$&$5.91_{-0.24}^{+0.09}$&$5.94_{-0.14}^{+0.02}$&$44.19_{-0.19}^{+0.60}$&$4.99_{-0.45}^{+0.01}$&$0.77_{-0.10}^{+0.40}$&$1.04_{-0.14}^{+0.27}$&$7.57_{-2.89}^{+0.36}$&428.91/356\\
193.19 $\sim$ 193.44 &$-0.62_{-0.05}^{+0.04}$& $-2.61_{-0.27}^{+0.14}$&$352.94_{-20.44}^{+26.65}$& 288.79/360&$3.00_{-0.21}^{+0.00}$&$5.83_{-0.35}^{+0.17}$&$5.80_{-0.07}^{+0.14}$&$44.04_{-0.04}^{+0.41}$&$4.99_{-0.54}^{+0.01}$&$1.27_{-0.39}^{+0.48}$&$1.05_{-0.15}^{+0.30}$&$6.57_{-0.49}^{+2.08}$&296.56/356\\
193.44 $\sim$ 194.14 &$-0.64_{-0.03}^{+0.02}$& $-2.51_{-0.13}^{+0.05}$&$447.21_{-12.38}^{+22.44}$& 436.49/360&$2.88_{-0.17}^{+0.12}$&$5.29_{-0.11}^{+0.45}$&$5.89_{-0.09}^{+0.07}$&$44.63_{-0.42}^{+0.51}$&$4.97_{-0.45}^{+0.02}$&$1.87_{-0.48}^{+0.48}$&$1.50_{-0.41}^{+0.07}$&$5.79_{-0.26}^{+2.52}$&443.12/356\\
194.14 $\sim$ 194.46 &$-0.66_{-0.05}^{+0.02}$& $-2.75_{-0.49}^{+0.12}$&$464.39_{-15.06}^{+41.72}$& 409.08/360&$2.96_{-0.17}^{+0.04}$&$5.97_{-0.27}^{+0.03}$&$5.81_{-0.06}^{+0.14}$&$45.81_{-0.55}^{+0.55}$&$4.47_{-0.73}^{+0.32}$&$1.79_{-0.86}^{+0.21}$&$1.44_{-0.36}^{+0.22}$&$3.50_{-0.74}^{+1.73}$&406.22/356\\
194.46 $\sim$ 194.60 &$-0.69_{-0.06}^{+0.02}$& $-2.85_{-10.72}^{+0.09}$&$561.04_{-23.25}^{+58.06}$& 350.66/360&$2.84_{-0.37}^{+0.04}$&$5.45_{-0.51}^{+0.29}$&$5.89_{-0.14}^{+0.06}$&$46.54_{-0.73}^{+0.32}$&$4.70_{-0.69}^{+0.29}$&$1.57_{-0.60}^{+0.39}$&$1.60_{-0.53}^{+0.10}$&$2.83_{-0.30}^{+2.14}$&350.11/356\\
194.60 $\sim$ 194.76 &$-0.66_{-0.07}^{+0.00}$& $-2.86_{-6.55}^{+0.48}$&$497.70_{-33.32}^{+33.32}$& 383.59/360&$2.64_{-0.23}^{+0.23}$&$5.96_{-0.28}^{+0.04}$&$5.97_{-0.10}^{+0.03}$&$45.32_{-0.98}^{+0.19}$&$4.65_{-1.26}^{+0.02}$&$1.18_{-0.21}^{+0.88}$&$1.45_{-0.21}^{+0.20}$&$5.42_{-0.58}^{+2.79}$&383.04/356\\
194.74 $\sim$ 195.06 &$-0.61_{-0.04}^{+0.02}$& $-2.54_{-0.15}^{+0.08}$&$428.72_{-15.98}^{+27.96}$& 398.31/360&$2.75_{-0.34}^{+0.05}$&$5.63_{-0.45}^{+0.20}$&$5.95_{-0.15}^{+0.01}$&$44.02_{-0.02}^{+0.54}$&$4.95_{-0.53}^{+0.04}$&$2.39_{-0.65}^{+0.60}$&$1.84_{-0.25}^{+0.16}$&$8.32_{-2.74}^{+0.08}$&408.56/356\\
195.06 $\sim$ 195.36 &$-0.66_{-0.03}^{+0.03}$& $-2.64_{-0.21}^{+0.12}$&$447.50_{-18.23}^{+30.07}$& 352.19/360&$2.56_{-0.11}^{+0.28}$&$5.58_{-0.53}^{+0.20}$&$5.89_{-0.11}^{+0.06}$&$44.84_{-0.52}^{+0.52}$&$4.77_{-0.50}^{+0.22}$&$0.86_{-0.19}^{+0.67}$&$1.18_{-0.01}^{+0.59}$&$6.37_{-0.73}^{+1.84}$&352.20/356\\
195.36 $\sim$ 195.66 &$-0.59_{-0.05}^{+0.04}$& $-2.39_{-0.14}^{+0.09}$&$383.95_{-21.04}^{+31.44}$& 363.67/360&$2.62_{-0.13}^{+0.23}$&$5.14_{-0.23}^{+0.53}$&$5.96_{-0.10}^{+0.04}$&$44.34_{-0.04}^{+1.13}$&$4.69_{-0.57}^{+0.30}$&$0.84_{-0.17}^{+0.49}$&$1.14_{-0.04}^{+0.39}$&$8.23_{-3.48}^{+0.24}$&370.09/356\\
195.66 $\sim$ 195.86 &$-0.66_{-0.04}^{+0.04}$& $-2.92_{-6.64}^{+0.58}$&$454.92_{-30.98}^{+30.98}$& 344.98/360&$2.84_{-0.37}^{+0.03}$&$5.86_{-0.53}^{+0.03}$&$5.95_{-0.18}^{+0.01}$&$44.29_{-0.04}^{+0.86}$&$4.93_{-0.59}^{+0.07}$&$1.47_{-0.50}^{+0.44}$&$1.33_{-0.21}^{+0.35}$&$7.18_{-1.56}^{+1.23}$&342.17/356\\
195.86 $\sim$ 196.16 &$-0.67_{-0.05}^{+0.03}$& $-2.84_{-10.43}^{+0.19}$&$440.17_{-20.83}^{+49.11}$& 405.11/360&$2.82_{-0.40}^{+0.02}$&$5.87_{-0.27}^{+0.12}$&$5.95_{-0.17}^{+0.01}$&$44.30_{-0.28}^{+0.52}$&$4.60_{-1.27}^{+0.01}$&$1.05_{-0.10}^{+0.96}$&$1.11_{-0.21}^{+0.29}$&$7.56_{-2.27}^{+1.09}$&401.11/356\\
196.16 $\sim$ 196.42 &$-0.66_{-0.07}^{+0.03}$& $-2.38_{-0.29}^{+0.07}$&$401.14_{-22.51}^{+53.30}$& 372.73/360&$2.69_{-0.28}^{+0.16}$&$4.74_{-0.20}^{+0.73}$&$5.98_{-0.10}^{+0.01}$&$44.88_{-0.52}^{+0.52}$&$4.66_{-0.71}^{+0.31}$&$1.69_{-0.74}^{+0.20}$&$1.72_{-0.31}^{+0.27}$&$6.56_{-0.87}^{+2.09}$&367.07/356\\
196.42 $\sim$ 196.82 &$-0.65_{-0.07}^{+0.00}$& $-2.76_{-11.85}^{+0.02}$&$394.09_{-27.97}^{+27.97}$& 414.59/360&$2.99_{-0.20}^{+0.01}$&$5.92_{-0.23}^{+0.08}$&$5.90_{-0.14}^{+0.05}$&$44.49_{-0.24}^{+0.70}$&$4.88_{-0.52}^{+0.12}$&$1.12_{-0.26}^{+0.41}$&$1.08_{-0.05}^{+0.36}$&$6.02_{-0.36}^{+2.11}$&413.25/356\\
196.82 $\sim$ 197.51 &$-0.73_{-0.04}^{+0.02}$& $-2.39_{-0.18}^{+0.06}$&$395.52_{-13.17}^{+38.33}$& 459.26/360&$2.64_{-0.29}^{+0.13}$&$4.51_{-0.19}^{+0.51}$&$5.81_{-0.16}^{+0.11}$&$44.05_{-0.05}^{+0.58}$&$4.21_{-1.54}^{+0.14}$&$1.18_{-0.31}^{+0.55}$&$1.18_{-0.08}^{+0.47}$&$8.88_{-1.42}^{+1.12}$&437.59/356\\
197.51 $\sim$ 198.07 &$-0.73_{-0.03}^{+0.01}$& $-3.12_{-6.21}^{+0.56}$&$538.38_{-4.14}^{+38.86}$& 435.96/360&$2.82_{-0.31}^{+0.02}$&$5.93_{-0.25}^{+0.07}$&$5.93_{-0.16}^{+0.02}$&$44.22_{-0.22}^{+0.47}$&$4.42_{-1.07}^{+0.20}$&$1.70_{-0.70}^{+0.35}$&$1.39_{-0.25}^{+0.33}$&$7.68_{-2.56}^{+0.38}$&424.35/356\\

\hline
 \multicolumn{14}{c}{ }\\
\end{tabular}
 \end{center}
\end{tiny}
\end{sidewaystable}

\setcounter{table}{0}

\begin{sidewaystable}
\setlength{\tabcolsep}{2pt}

\begin{tiny}
\begin{center}
\caption{Time-resolved Spectral Fitting (Sub-burst II, continued) }

\begin{tabular}{l|llll|lllllllll}
\hline
 \multicolumn{1}{c|}{Band Function Fitting} & \multicolumn{4}{c|}{Band Function Fitting} & \multicolumn{9}{c}{Synchrotron Model Fitting} \\
\hline

 t1 $\sim$ t2 (s) & $\alpha$ & $\beta$ & $E_p$ (keV) & PGSTAT/dof & log $\Gamma$ & $p$ & log $\gamma_{inj}$ & log R$_{inj}^0$ (s$^{-1}$) & q & log B$_0$ (G) & b & $\hat{t}$ (s) & PGSTAT/dof \\ 
\hline
\hline
198.07 $\sim$ 198.47 &$-0.71_{-0.04}^{+0.03}$& $-2.43_{-0.22}^{+0.08}$&$425.62_{-21.36}^{+39.95}$& 366.28/360&$2.65_{-0.20}^{+0.17}$&$4.50_{-0.06}^{+0.73}$&$5.91_{-0.21}^{+0.02}$&$44.16_{-0.16}^{+0.67}$&$4.90_{-0.74}^{+0.10}$&$1.06_{-0.10}^{+1.06}$&$1.25_{-0.10}^{+0.51}$&$7.91_{-3.44}^{+0.67}$&357.99/356\\
198.47 $\sim$ 198.83 &$-0.81_{-0.03}^{+0.03}$& $-3.54_{-6.07}^{+1.12}$&$483.81_{-24.78}^{+30.74}$& 353.60/360&$2.19_{-0.12}^{+0.25}$&$5.55_{-0.41}^{+0.30}$&$5.76_{-0.11}^{+0.10}$&$44.00_{-0.00}^{+0.51}$&$4.31_{-0.79}^{+0.62}$&$0.78_{-0.11}^{+0.60}$&$1.11_{-0.20}^{+0.29}$&$9.35_{-1.68}^{+0.64}$&346.85/356\\
198.83 $\sim$ 199.46 &$-0.74_{-0.03}^{+0.01}$& $-3.14_{-6.39}^{+0.50}$&$503.07_{-6.64}^{+30.23}$& 399.54/360&$2.66_{-0.22}^{+0.13}$&$5.98_{-0.25}^{+0.02}$&$5.91_{-0.16}^{+0.05}$&$44.08_{-0.08}^{+0.50}$&$4.51_{-0.92}^{+0.11}$&$1.59_{-0.65}^{+0.57}$&$1.45_{-0.33}^{+0.32}$&$8.47_{-3.56}^{+0.04}$&386.70/356\\
199.46 $\sim$ 199.89 &$-0.66_{-0.03}^{+0.02}$& $-2.79_{-0.30}^{+0.14}$&$578.36_{-23.07}^{+36.16}$& 389.83/360&$2.82_{-0.32}^{+0.06}$&$6.00_{-0.25}^{+0.00}$&$5.85_{-0.02}^{+0.12}$&$44.61_{-0.38}^{+0.49}$&$4.90_{-0.46}^{+0.10}$&$0.96_{-0.29}^{+0.51}$&$0.97_{-0.07}^{+0.32}$&$5.50_{-0.16}^{+2.56}$&392.21/356\\
199.89 $\sim$ 200.29 &$-0.68_{-0.03}^{+0.02}$& $-2.83_{-0.34}^{+0.13}$&$610.44_{-19.91}^{+42.63}$& 408.52/360&$2.93_{-0.15}^{+0.07}$&$6.00_{-0.23}^{+0.00}$&$5.93_{-0.11}^{+0.04}$&$44.71_{-0.51}^{+0.16}$&$4.61_{-0.75}^{+0.19}$&$1.34_{-0.38}^{+0.74}$&$1.16_{-0.16}^{+0.44}$&$5.77_{-1.01}^{+1.01}$&400.90/356\\
200.29 $\sim$ 200.38 &$-0.63_{-0.11}^{+0.04}$& $-2.27_{-0.41}^{+0.09}$&$532.26_{-37.85}^{+155.08}$& 412.93/360&$2.68_{-0.24}^{+0.17}$&$4.36_{-0.29}^{+0.78}$&$5.96_{-0.10}^{+0.04}$&$44.53_{-0.23}^{+1.08}$&$3.65_{-0.98}^{+0.70}$&$0.69_{-0.01}^{+0.68}$&$0.92_{-0.02}^{+0.29}$&$9.09_{-1.96}^{+0.88}$&408.68/356\\
200.38 $\sim$ 200.48 &$-0.84_{-0.04}^{+0.03}$& $-2.66_{-11.19}^{+0.06}$&$840.16_{-62.07}^{+110.54}$& 386.58/360&$2.18_{-0.14}^{+0.23}$&$4.12_{-0.27}^{+1.02}$&$5.80_{-0.18}^{+0.09}$&$44.61_{-0.13}^{+1.19}$&$4.00_{-2.23}^{+0.13}$&$0.88_{-0.21}^{+0.64}$&$1.12_{-0.05}^{+0.61}$&$8.09_{-3.48}^{+0.37}$&379.27/356\\
200.48 $\sim$ 200.68 &$-0.68_{-0.04}^{+0.04}$& $-1.96_{-0.05}^{+0.06}$&$667.35_{-65.31}^{+72.39}$& 436.58/360&$2.92_{-0.21}^{+0.08}$&$3.14_{-0.15}^{+0.08}$&$5.90_{-0.19}^{+0.02}$&$44.24_{-0.23}^{+0.53}$&$4.92_{-0.74}^{+0.08}$&$2.69_{-0.65}^{+0.30}$&$1.78_{-0.34}^{+0.22}$&$7.00_{-2.04}^{+1.47}$&428.06/356\\
200.68 $\sim$ 200.98 &$-0.63_{-0.03}^{+0.03}$& $-2.12_{-0.06}^{+0.05}$&$600.97_{-33.09}^{+39.29}$& 434.72/360&$2.83_{-0.27}^{+0.07}$&$3.63_{-0.10}^{+0.18}$&$5.99_{-0.08}^{+0.01}$&$45.31_{-1.02}^{+0.01}$&$4.72_{-0.85}^{+0.10}$&$1.07_{-0.10}^{+1.05}$&$1.23_{-0.14}^{+0.46}$&$5.12_{-0.00}^{+2.74}$&452.01/356\\
200.98 $\sim$ 201.23 &$-0.71_{-0.03}^{+0.03}$& $-2.53_{-0.19}^{+0.11}$&$606.95_{-32.58}^{+46.84}$& 345.99/360&$2.81_{-0.36}^{+0.05}$&$4.68_{-0.18}^{+0.68}$&$5.81_{-0.04}^{+0.15}$&$45.17_{-0.74}^{+0.38}$&$4.30_{-1.20}^{+0.19}$&$0.90_{-0.23}^{+0.55}$&$0.95_{-0.05}^{+0.29}$&$4.90_{-1.49}^{+1.49}$&339.38/356\\
201.23 $\sim$ 201.54 &$-0.51_{-0.06}^{+0.04}$& $-2.25_{-0.14}^{+0.05}$&$348.59_{-20.60}^{+32.31}$& 353.86/360&$2.84_{-0.34}^{+0.00}$&$5.14_{-0.23}^{+0.60}$&$5.93_{-0.13}^{+0.03}$&$45.14_{-0.86}^{+0.19}$&$4.86_{-0.66}^{+0.14}$&$1.77_{-0.88}^{+0.17}$&$1.59_{-0.49}^{+0.05}$&$5.09_{-1.35}^{+1.35}$&372.07/356\\
201.54 $\sim$ 201.83 &$-0.77_{-0.03}^{+0.02}$& $-3.69_{-5.94}^{+1.19}$&$549.81_{-16.97}^{+34.85}$& 427.24/360&$2.76_{-0.44}^{+0.01}$&$5.87_{-0.31}^{+0.13}$&$5.65_{-0.11}^{+0.11}$&$44.58_{-0.22}^{+0.82}$&$3.72_{-0.78}^{+0.73}$&$1.43_{-0.50}^{+0.52}$&$1.05_{-0.15}^{+0.32}$&$5.99_{-0.50}^{+2.69}$&415.51/356\\
201.83 $\sim$ 202.10 &$-0.67_{-0.05}^{+0.00}$& $-3.15_{-6.53}^{+0.60}$&$547.66_{-25.90}^{+25.90}$& 437.87/360&$2.63_{-0.16}^{+0.20}$&$5.91_{-0.24}^{+0.09}$&$5.96_{-0.08}^{+0.04}$&$44.11_{-0.11}^{+0.53}$&$4.97_{-0.52}^{+0.02}$&$0.70_{-0.03}^{+0.44}$&$0.99_{-0.07}^{+0.26}$&$7.96_{-3.02}^{+0.04}$&436.43/356\\
202.10 $\sim$ 202.38 &$-0.69_{-0.03}^{+0.03}$& $-2.88_{-6.54}^{+0.44}$&$552.61_{-35.35}^{+35.35}$& 413.69/360&$2.88_{-0.20}^{+0.12}$&$5.92_{-0.27}^{+0.08}$&$5.99_{-0.09}^{+0.01}$&$44.40_{-0.02}^{+1.30}$&$4.42_{-0.87}^{+0.27}$&$1.22_{-0.31}^{+0.57}$&$1.16_{-0.08}^{+0.43}$&$7.42_{-3.11}^{+0.34}$&406.08/356\\
202.38 $\sim$ 202.68 &$-0.81_{-0.03}^{+0.03}$& $-18.08_{-1.92}^{+5.66}$&$464.92_{-22.13}^{+23.30}$& 400.62/360&$2.66_{-0.34}^{+0.08}$&$5.98_{-0.33}^{+0.02}$&$5.67_{-0.03}^{+0.24}$&$44.88_{-0.47}^{+0.65}$&$3.33_{-0.60}^{+1.12}$&$1.42_{-0.44}^{+0.71}$&$1.19_{-0.27}^{+0.30}$&$6.14_{-0.79}^{+2.08}$&393.52/356\\
202.68 $\sim$ 203.03 &$-0.78_{-0.05}^{+0.01}$& $-2.86_{-11.97}^{+0.05}$&$482.36_{-7.62}^{+55.58}$& 411.44/360&$2.82_{-0.23}^{+0.17}$&$5.45_{-0.53}^{+0.30}$&$5.79_{-0.12}^{+0.12}$&$44.68_{-0.37}^{+0.75}$&$2.97_{-0.34}^{+1.25}$&$1.33_{-0.26}^{+0.90}$&$1.10_{-0.16}^{+0.25}$&$7.69_{-2.17}^{+1.10}$&407.68/356\\
203.03 $\sim$ 203.61 &$-0.78_{-0.04}^{+0.01}$& $-3.12_{-6.51}^{+0.71}$&$453.85_{-19.98}^{+19.98}$& 426.54/360&$2.41_{-0.08}^{+0.41}$&$5.85_{-0.38}^{+0.15}$&$5.82_{-0.14}^{+0.10}$&$44.52_{-0.24}^{+0.76}$&$3.47_{-0.58}^{+0.85}$&$0.97_{-0.03}^{+1.32}$&$1.18_{-0.01}^{+0.63}$&$9.02_{-1.62}^{+0.97}$&418.06/356\\
203.61 $\sim$ 204.21 &$-0.90_{-0.03}^{+0.02}$& $-3.11_{-6.33}^{+0.70}$&$457.73_{-6.54}^{+44.03}$& 389.27/360&$2.03_{-0.03}^{+0.24}$&$5.96_{-0.54}^{+0.04}$&$5.73_{-0.19}^{+0.08}$&$44.61_{-0.14}^{+1.22}$&$3.24_{-2.30}^{+0.07}$&$0.92_{-0.25}^{+0.58}$&$1.43_{-0.23}^{+0.29}$&$9.23_{-1.79}^{+0.75}$&384.33/356\\
204.21 $\sim$ 205.21 &$-0.81_{-0.04}^{+0.02}$& $-2.58_{-0.64}^{+0.07}$&$433.79_{-9.55}^{+48.06}$& 431.95/360&$2.66_{-0.25}^{+0.16}$&$4.53_{-0.17}^{+0.62}$&$5.48_{-0.16}^{+0.16}$&$44.59_{-0.20}^{+0.95}$&$3.93_{-1.72}^{+0.29}$&$1.60_{-0.53}^{+0.57}$&$1.15_{-0.02}^{+0.54}$&$4.91_{-0.68}^{+2.57}$&415.16/356\\
205.21 $\sim$ 205.78 &$-0.74_{-0.04}^{+0.04}$& $-2.74_{-6.81}^{+0.51}$&$382.07_{-26.41}^{+26.41}$& 378.63/360&$2.58_{-0.11}^{+0.26}$&$5.52_{-0.58}^{+0.23}$&$5.74_{-0.08}^{+0.18}$&$44.55_{-0.18}^{+0.77}$&$3.86_{-1.11}^{+0.41}$&$0.85_{-0.17}^{+0.55}$&$0.99_{-0.09}^{+0.31}$&$6.81_{-1.62}^{+1.46}$&372.13/356\\
205.78 $\sim$ 206.18 &$-0.79_{-0.05}^{+0.02}$& $-3.27_{-6.46}^{+0.93}$&$449.60_{-13.13}^{+43.15}$& 367.42/360&$2.44_{-0.05}^{+0.36}$&$5.59_{-0.66}^{+0.18}$&$5.66_{-0.03}^{+0.22}$&$44.14_{-0.13}^{+0.63}$&$4.51_{-0.98}^{+0.47}$&$1.09_{-0.15}^{+0.90}$&$1.11_{-0.21}^{+0.27}$&$6.90_{-2.62}^{+1.37}$&363.85/356\\
206.18 $\sim$ 206.78 &$-0.69_{-0.08}^{+0.01}$& $-2.66_{-5.76}^{+0.38}$&$258.20_{-0.43}^{+33.32}$& 430.15/360&$2.68_{-0.25}^{+0.19}$&$5.95_{-0.36}^{+0.05}$&$5.81_{-0.18}^{+0.10}$&$44.36_{-0.18}^{+0.73}$&$4.54_{-0.65}^{+0.45}$&$1.08_{-0.16}^{+0.98}$&$1.25_{-0.04}^{+0.44}$&$6.83_{-1.29}^{+1.69}$&434.20/356\\
206.78 $\sim$ 207.78 &$-0.75_{-0.06}^{+0.03}$& $-2.31_{-0.24}^{+0.07}$&$256.48_{-10.45}^{+30.57}$& 433.43/360&$2.71_{-0.35}^{+0.01}$&$4.39_{-0.21}^{+0.64}$&$5.91_{-0.15}^{+0.09}$&$44.71_{-0.37}^{+0.68}$&$4.38_{-0.86}^{+0.61}$&$0.90_{-0.02}^{+0.90}$&$1.30_{-0.11}^{+0.42}$&$6.61_{-2.09}^{+1.66}$&433.21/356\\
207.78 $\sim$ 208.37 &$-0.77_{-0.11}^{+0.06}$& $-2.38_{-2.31}^{+0.13}$&$220.82_{-15.38}^{+44.47}$& 394.46/360&$2.51_{-0.14}^{+0.32}$&$4.18_{-0.17}^{+0.96}$&$5.86_{-0.18}^{+0.13}$&$45.40_{-0.93}^{+0.23}$&$3.96_{-1.00}^{+1.00}$&$1.10_{-0.20}^{+0.77}$&$1.79_{-0.31}^{+0.21}$&$5.52_{-0.73}^{+2.09}$&395.14/356\\
208.37 $\sim$ 208.97 &$-0.88_{-0.08}^{+0.03}$& $-3.08_{-6.10}^{+0.79}$&$236.63_{-10.57}^{+24.89}$& 387.04/360&$2.73_{-0.22}^{+0.22}$&$5.48_{-0.79}^{+0.27}$&$5.37_{-0.16}^{+0.16}$&$45.83_{-0.66}^{+0.66}$&$1.82_{-0.46}^{+1.54}$&$2.01_{-0.51}^{+0.51}$&$1.72_{-0.53}^{+0.07}$&$2.27_{-0.64}^{+2.40}$&388.02/356\\
208.97 $\sim$ 210.47 &$-0.87_{-0.08}^{+0.01}$& $-2.56_{-6.82}^{+0.43}$&$251.89_{-18.88}^{+18.88}$& 376.01/360&$2.88_{-0.24}^{+0.12}$&$4.59_{-0.44}^{+0.62}$&$5.50_{-0.10}^{+0.20}$&$45.09_{-0.74}^{+0.15}$&$1.90_{-0.36}^{+1.37}$&$1.12_{-0.18}^{+1.04}$&$0.95_{-0.04}^{+0.35}$&$4.32_{-1.87}^{+1.87}$&372.58/356\\
210.47 $\sim$ 211.81 &$-0.88_{-0.09}^{+0.01}$& $-2.46_{-6.72}^{+0.42}$&$276.17_{-25.36}^{+25.36}$& 421.23/360&$2.41_{-0.12}^{+0.39}$&$4.06_{-0.03}^{+1.11}$&$5.57_{-0.09}^{+0.27}$&$44.86_{-0.32}^{+0.93}$&$2.47_{-1.61}^{+0.42}$&$0.96_{-0.02}^{+1.29}$&$1.13_{-0.21}^{+0.26}$&$7.05_{-1.90}^{+1.48}$&415.96/356\\
211.81 $\sim$ 214.97 &$-1.06_{-0.07}^{+0.07}$& $-2.13_{-6.53}^{+0.23}$&$178.72_{-34.12}^{+34.12}$& 404.18/360&$2.50_{-0.36}^{+0.11}$&$3.20_{-0.19}^{+1.35}$&$5.19_{-0.08}^{+0.20}$&$44.60_{-0.18}^{+0.93}$&$0.95_{-0.61}^{+0.99}$&$2.04_{-0.88}^{+0.28}$&$1.36_{-0.18}^{+0.46}$&$6.65_{-3.55}^{+1.31}$&408.33/356\\
214.97 $\sim$ 215.47 &$-1.15_{-0.26}^{+0.10}$& $-3.79_{-6.12}^{+1.92}$&$255.96_{-28.16}^{+581.04}$& 281.59/360&$2.90_{-0.27}^{+0.10}$&$4.89_{-1.61}^{+0.41}$&$4.93_{-0.20}^{+0.20}$&$44.03_{-0.02}^{+0.45}$&$2.38_{-1.65}^{+0.29}$&$2.77_{-0.63}^{+0.22}$&$1.22_{-0.11}^{+0.51}$&$2.56_{-2.21}^{+2.52}$&281.33/356\\
215.47 $\sim$ 218.24 &$-1.16_{-0.09}^{+0.09}$& $-1.96_{-6.95}^{+0.32}$&$198.34_{-70.43}^{+99.97}$& 363.97/360&$2.19_{-0.06}^{+0.44}$&$2.42_{-0.03}^{+1.77}$&$5.18_{-0.10}^{+0.19}$&$45.13_{-0.41}^{+0.41}$&$0.48_{-0.11}^{+1.48}$&$1.16_{-0.47}^{+0.58}$&$1.25_{-0.13}^{+0.51}$&$4.28_{-1.25}^{+3.90}$&365.23/356\\

\hline
\end{tabular}
 \end{center}
\end{tiny}
\end{sidewaystable}

\begin{table*}

\caption{Derived Photoshere Radii and Lorentz Factors for Sub-burst I}

\label{tab:physicalpar}
\begin{tiny}
\begin{tabular}{c|cccc}
\hline

(t$_1$,t$_2$) & $r_{0}\times Y^{3/2}$ (cm) &$ \Gamma \times Y^{-1/4}$& & PGSTAT/dof\\
\hline
(-0.30,-0.08)& $4.6\times 10^{8}$ & 214 & &203.42/362\\
(-0.08,0.13)& $4.6\times 10^{8}$ & 323 & &222.68/362\\
(0.13,0.35)& $5.2\times 10^{8}$ & 381 & &278.12/362\\
(0.35,0.57)& $4.9\times 10^{8}$& 370 & &228.15/362\\
(0.57,0.78)& $4.7\times 10^{8}$ &348 & &209.76/362\\
(0.78,1.00)& $7.2\times 10^{8}$& 224 & &217.85/362\\
\hline

\end{tabular}
\end{tiny}
\end{table*}

\newpage
\clearpage

\begin{center}
\footnotesize
\begin{longtable*}{|cccccccccc|}

 \caption{Optical observations of GRB 160625B.}\\
\hline
 $T_{\rm mid}$(s) & $T_{\rm exp}$(s) & Mag & Error & Band & $T_{\rm mid}$(s) & $T_{\rm exp}$(s) & Mag & Error & Band\\
\hline
\endfirsthead
\multicolumn{10}{c}%
{ -- \textit{Continued from previous page}} \\
\hline
 $T_{mid}$(s) & $T_{exp}$(s) & Mag & Error & Band & $T_{mid}$(s) & $T_{exp}$(s) & Mag & Error & Band\\
\hline
\endhead
\hline 
\multicolumn{10}{r}{\textit{Continued on next page}} \\
\endfoot
\hline
\endlastfoot

\multicolumn{10}{|c|}{BOOTES-1 and BOOTES-2 }\\
\hline
-27.28 & 36 & $>$7.5 & & R & 41.72 & $>$ 36 & $>$7.5 & & R \\
\hline
\multicolumn{10}{|c|}{Mini-MegaTORTORA }\\
\hline
187.90 & 30 & 8.67 & 0.004 & V & 429.30 & 30 & 11.16 & 0.02 & V \\
218.00 & 30 & 8.52 & 0.004 & V & 459.40 & 30 & 11.28 & 0.02 & V \\
248.10 & 30 & 9.19 & 0.01 & V & 489.50 & 30 & 11.50 & 0.03 & V \\
278.20 & 30 & 9.53 & 0.01 & V & 519.60 & 30 & 11.63 & 0.03 & V \\
308.60 & 30 & 9.91 & 0.01 & V & 550.00 & 30 & 11.78 & 0.03 & V \\
338.70 & 30 & 10.24 & 0.01 & V & 580.10 & 30 & 11.87 & 0.04 & V \\
368.80 & 30 & 10.56 & 0.01 & V & 610.20 & 30 & 11.98 & 0.04 & V \\
398.90 & 30 & 10.85 & 0.02 & V & 640.30 & 30 & 12.14 & 0.05 & V \\

67.20 & 30 & $>$13.85 & & V & 2136.60 & 30 & $>$13.82 & & V \\
97.30 & 30 & $>$13.76 & & V & 2166.69 & 30 & $>$13.68 & & V \\
127.40 & 30 & $>$13.75 & & V & 2196.79 & 30 & $>$13.78 & & V \\
157.50 & 30 & $>$13.9 & & V & 2226.89 & 30 & $>$13.83 & & V \\
1895.19 & 30 & $>$13.79 & & V & 2257.30 & 30 & $>$13.8 & & V \\
1925.29 & 30 & $>$13.74 & & V & 2287.39 & 30 & $>$13.84 & & V \\
1955.39 & 30 & $>$13.7 & & V & 2317.40 & 30 & $>$13.79 & & V \\
1985.49 & 30 & $>$13.79 & & V & 2347.49 & 30 & $>$13.82 & & V \\
2015.89 & 30 & $>$13.67 & & V & 2377.89 & 30 & $>$13.7 & & V \\
2045.99 & 30 & $>$13.71 & & V & 2407.90 & 30 & $>$13.66 & & V \\
2076.09 & 30 & $>$13.8 & & V & 2437.99 & 30 & $>$13.7 & & V \\
2106.19 & 30 & $>$13.76 & & V & 2468.09 & 30 & $>$13.73 & & V \\
\hline
\multicolumn{10}{|c|}{Pi of the Sky 35 }\\
\hline
187.10 & 10 & 9.18 & 0.02 & V & 369.79 & 10 & 11.11 & 0.12 & V \\
200.30 & 10 & 8.04 & 0.01 & V & 383.00 & 10 & 11.28 & 0.14 & V \\
213.50 & 10 & 8.85 & 0.01 & V & 396.20 & 10 & 11.43 & 0.16 & V \\
226.71 & 10 & 9.16 & 0.02 & V & 409.41 & 10 & 11.52 & 0.17 & V \\
239.91 & 10 & 9.35 & 0.02 & V & 422.60 & 10 & 11.66 & 0.20 & V \\
253.13 & 10 & 9.48 & 0.03 & V & 435.80 & 10 & 11.97 & 0.26 & V \\
266.33 & 10 & 9.76 & 0.03 & V & 448.99 & 10 & 11.44 & 0.16 & V \\
279.53 & 10 & 9.89 & 0.04 & V & 462.19 & 10 & 12.23 & 0.33 & V \\
303.76 & 10 & 10.36 & 0.06 & V & 475.39 & 10 & 11.71 & 0.20 & V \\
316.97 & 10 & 10.36 & 0.06 & V & 488.60 & 10 & 12.10 & 0.30 & V \\
330.17 & 10 & 10.66 & 0.08 & V & 501.80 & 10 & 12.02 & 0.27 & V \\
343.38 & 10 & 10.78 & 0.09 & V & 541.41 & 10 & 12.15 & 0.30 & V \\
356.58 & 10 & 10.90 & 0.10 & V & 581.02 & 10 & 12.21 & 0.33 & V \\

\hline
\multicolumn{10}{|c|}{Pi of the Sky 39 }\\
\hline
192.78 & 10 & 8.08 & 0.01 & V & 362.12 & 10 & 10.99 & 0.12 & V \\
208.13 & 10 & 8.40 & 0.01 & V & 377.47 & 10 & 11.23 & 0.15 & V \\
223.48 & 10 & 9.30 & 0.03 & V & 392.82 & 10 & 11.05 & 0.13 & V \\
238.83 & 10 & 9.38 & 0.03 & V & 408.17 & 10 & 11.27 & 0.16 & V \\
254.69 & 10 & 9.60 & 0.03 & V & 425.57 & 10 & 11.47 & 0.19 & V \\
270.04 & 10 & 9.76 & 0.04 & V & 440.89 & 10 & 12.00 & 0.31 & V \\
285.39 & 10 & 10.00 & 0.05 & V & 456.75 & 10 & 11.79 & 0.25 & V \\
300.72 & 10 & 10.24 & 0.06 & V & 472.10 & 10 & 12.09 & 0.34 & V \\
316.07 & 10 & 10.44 & 0.07 & V & 487.45 & 10 & 11.95 & 0.30 & V \\
331.42 & 10 & 10.64 & 0.09 & V & 518.15 & 10 & 12.14 & 0.35 & V \\
346.77 & 10 & 10.79 & 0.10 & V & 625.60 & 10 & 12.19 & 0.37 & V \\

\hline
\multicolumn{10}{|c|}{TSHAO }\\
\hline 
79301.38 & 6960 & 18.90 & 0.05 & R & 2405985.98 & 6600 & $>$21.7 & & R \\
247220.64 & 7560 & 20.32 & 0.06 & R & 2491992.00 & 6600 & $>$22.0 & & R \\
764124.19 & 2400 & 21.33 & 0.18 & R & 3182164.70 & 3120 & $>$21.9 & & R \\
1029271.10 & 2160 & 21.67 & 0.29 & R & 3273906.82 & 10080 & $>$22.7 & & R \\
1116053.86 & 11700 & 21.61 & 0.16 & R & 3354166.37 & 5040 & $>$22.5 & & R \\
1284887.23 & 8700 & 21.64 & 0.09 & R & 3268571.62 & 18240 & $>$23.0 & & R \\
1378602.72 & 5100 & 21.89 & 0.34 & R & 3785611.68 & 2760 & $>$22.0 & & R \\
1464446.30 & 5700 & 21.92 & 0.28 & R & 3964933.15 & 1560 & $>$21.1 & & R \\

\hline
\multicolumn{10}{|c|}{AbAO }\\
\hline
166746.82 & 3240 & 19.61 & 0.06 & CR & 1731695.33 & 4020 & 22.13 & 0.21 & CR \\
424964.45 & 1380 & 20.75 & 0.37 & CR & 1896639.84 & 3000 & $>$21.7 & & CR \\
518178.82 & 1980 & 20.89 & 0.17 & CR & 2594973.89 & 3840 & 22.24 & 0.26 & CR \\
693245.09 & 1620 & 21.22 & 0.18 & CR & 2682638.78 & 3840 & $>$22.5 & & CR \\
1559670.34 & 2880 & 21.87 & 0.35 & CR & 3026274.05 & 4800 & $>$22.9 & & CR \\
\hline
\multicolumn{10}{|c|}{Mondy }\\
\hline
328701.02 & 3600 & 20.33 & 0.10 & R & 1360882.94 & 3600 & 21.93 & 0.11 & R \\
757106.78 & 3480 & 21.13 & 0.09 & R & 2394252.00 & 4500 & $>$21.6 & & R \\
\hline
\multicolumn{10}{|c|}{CrAO }\\
\hline
84461.18 & 3240 & 18.87 & 0.10 & R & 2935651.68 & 3600 & 23.05 & 0.08 & R \\
2680685.28 & 4080 & 22.88 & 0.09 & R & 5730540.48 & 3840 & $>$23.8 & & R \\
2847137.47 & 5760 & 23.10 & 0.07 & R &&&&&\\\hline
\multicolumn{10}{|c|}{Maidanak }\\
\hline
1122346.37 & 1800 & 21.75 & 0.11 & R & 1891335.74 & 5400 & 22.30 & 0.34 & R \\
1208215.01 & 1800 & 21.55 & 0.08 & R & 2326797.79 & 4320 & 22.45 & 0.26 & R \\
1294950.24 & 1800 & 21.71 & 0.09 & R & 3480784.70 & 3600 & 23.31 & 0.10 & R \\
1373255.42 & 1800 & 21.97 & 0.11 & R & 3830392.80 & 13800 & 23.69 & 0.11 & R \\
1462217.18 & 1800 & 22.02 & 0.10 & R & 4608813.6 & 8400 & $>$22.8 & & R \\
1552235.62 & 1800 & 22.12 & 0.17 & R &&&&& \\

\hline
%
\multicolumn{10}{|c|}{SAO RAS~}\\
\hline
170726.40 & 2100 & 19.87 & 0.07 & R & 348710.40 & 2400 & 20.56 & 0.20 & R \\
335232.00 & 600 & 20.53 & 0.07 & R & 3276451.30 & 720 & 23.47 & 0.10 & R \\
\hline

\label{tab:optobs}
\end{longtable*}
\end{center}

\newpage 
\begin{table*}
\caption{Time lag of optical light curves with respect to the $\gamma$-ray light curve in the 2nd sub-burst}
\begin{tabular}{cc}
\hline
Band & lag (s) \\
\hline
Pi 35 vs GBM & 3.1$_{-2.1}^{+1.4}$ \\ 
Pi 39 vs GBM & 3.3$_ {-1.1}^{ +1.3}$ \\ 
Mini-MegaTORTORA vs GBM & 4.8$_{-4.7}^{+5.1}$\\
\hline
\end{tabular}
\label{tab:lag_opt_gbm}
\end{table*}

\begin{table*}

\caption{Spectral fitting parameters for the three SED slices}
\begin{tabular}{c|ccccc}

\hline

\hline
(t$_1$,t$_2$) &kT (keV) & $\alpha$ & $\beta$ & E$_p$ (keV) & PGSTAT/dof\\
\hline

(172.90,202.90) & & $-0.65_{-0.01}^{+0.01}$& $-2.18_{-0.01}^{+0.01}$& $466.80_{-4.63}^{+5.28}$ &3799.78/360\\
(203.00,233.00) & & $-0.85_{-0.01}^{+0.09}$& $-2.21_{-0.05}^{+0.03}$& $302.63_{-6.83}^{+8.61}$ &1724.41/360\\
(233.10,263.10) & & $-0.54_{-0.23}^{+0.15}$& $-1.70_{-0.24}^{+0.05}$& $454.01_{-81.78}^{+271.34}$ &407.87/360\\
(172.90,202.90) & $35.00_{-0.56}^{+0.55}$& $-0.79_{-0.01}^{+0.01}$& $-2.40_{-0.01}^{+0.02}$& $744.88_{-13.59}^{+11.34}$ &2745.70/360\\
(203.00,233.00) & $33.22_{-1.10}^{+1.31}$& $-1.06_{-0.02}^{+0.02}$& $-2.43_{-0.11}^{+0.08}$& $493.80_{-26.01}^{+29.46}$ &1655.76/360\\
\hline

\end{tabular}

\label{tab:sed}

\end{table*}

\makeatletter
\apptocmd{\thebibliography}{\global\c@NAT@ctr 47\relax}{}{}

\makeatother

\makeatletter
\makeatother

\newpage
\makeatletter
\def\@biblabel#1{\@ifnotempty{#1}{#1.}}
\makeatother

\end{document}